   \def\pfeil{_\rightharpoonup}   \def\leer{\phantom{a}}
   \def\opf{\buildrel \pfeil \over \leer}
   \def\jvv{j \lower0.4pt\hbox to 2pt{\hss $\opf$}}

   \def\jv{j \lower0.2pt\hbox to 1.4pt{\hss $\opf$}}
   \def\ivv{i \lower0.4pt\hbox to 2pt{\hss $\opf$}}
   \def\iv{i \lower0.2pt\hbox to 1.4pt{\hss $\opf$}}
   \def\hq{h \raise0.2pt\hbox to 0.4pt{\hss $^-$}}
   \def\vk#1{\hbox{$\buildrel           \pfeil \over #1$}}
   \def\vkk#1{\hbox{$\buildrel   \;     \pfeil \over #1$}}
   \def\vkkk#1{\hbox{$\buildrel  \, \;  \pfeil \over #1$}}
   \def\grpf{\displaystyle  _\rightharpoonup}
   \def\vg#1{\hbox{$\buildrel       \grpf \over #1$}}
   \def\vgg#1{\hbox{$\buildrel  \;  \grpf \over #1$}}
\def\fzz{f} \def\bzz{b} \def\dzz{d} \def\gzz{g} \def\hzz{h} \def\jzz{j}
\def\kzz{k} \def\lzz{l} \def\mzz{m} \def\wzz{w} \def\tzz{t} \def\izz{i}
\def\bezz{\beta} \def\dezz{\delta} \def\xizz{\xi} \def\pszz{\psi}
\def\vthzz{\vartheta}
\def\uph{ \! \mathop{\vphantom{a}} } \def\dph{ \vphantom{a} }
\def\vc#1{\def\tast{\noexpand#1} \def\test{#1}
    \ifcat\tast\bzz
      \ifx\test\fzz \vkkk f \uph \else
       \ifx\test\bzz \vkk b \uph \else
        \ifx\test\dzz \vkkk d \uph \else
         \ifx\test\gzz \vkk g \dph \else
          \ifx\test\hzz \vkk h \uph \else
           \ifx\test\izz \ivv \dph \else
            \ifx\test\jzz \jvv \dph \else
             \ifx\test\kzz \vkk k \uph \else
              \ifx\test\lzz \vkk l \uph \else
               \ifx\test\tzz \vkk t \uph \else
                \ifx\test\mzz \vg m \dph \else
                 \ifx\test\wzz \vg w \dph \else
                  \ifnum \lq#1<91 \vgg #1 \uph \else \vk #1 \dph
                  \fi
                 \fi
                \fi
               \fi
              \fi
             \fi
            \fi
           \fi
          \fi
         \fi
        \fi
       \fi
      \fi
    \else
     \ifx\test\bezz \vkk \beta \uph \else
      \ifx\test\pszz \vkk \psi \dph \else
       \ifx\test\dezz \vkk \delta \uph \else
        \ifx\test\xizz \vkk \xi \uph \else
         \ifx\test\vthzz \vkk \vartheta \uph \else \vk #1 \dph
         \fi
        \fi
       \fi
      \fi
     \fi
    \fi }        
\def\fgra{                      
\def\d{{$\!$.}}
\begin{figure}[t] \unitlength.9cm \begin{picture}(15.2,6)
\put(3,.5){\vector(1,0){12.4}}   \put(15.6,.45){$g$}
\put(3,.5){\vector(0,1){4.7}}  \put(2.94,5.4){$\omega$}
\thicklines \put(3,.5){\line(2,3){3.2}} \thinlines
      \put(5,3.2){\oval(2.6,3.6)[bl]}
      \put(5,3.2){\oval(2.6,3.6)[br]}
      \put(5,3.2){\oval(2.6,3.6)[tl]}
      \put(5,3.2){\oval(2.6,3.6)[tr]}
\put(9.3,4){\frm pure}  \put(8.8,3.4){\frm speculation}
\put(12.5,1){\frm glue-ball gas ?}
\put(6.1,5.57){$m$}
\thicklines \put(11.98,.38){\line(0,1){.24}} \thinlines
\def\dd{{\circle*{.06}}}
\put(3.9162,1.8024)\dd \put(4.0357,1.9615)\dd \put(4.1565,2.1195)\dd
\put(4.2787,2.2765)\dd \put(4.4021,2.4324)\dd \put(4.5270,2.5872)\dd
\put(4.6533,2.7407)\dd \put(4.7812,2.8928)\dd \put(4.9107,3.0436)\dd
\put(5.0418,3.1929)\dd \put(5.1746,3.3406)\dd \put(5.3093,3.4866)\dd
\put(5.4458,3.6308)\dd \put(5.5844,3.7730)\dd \put(5.7250,3.9130)\dd
\put(5.8678,4.0508)\dd \put(6.0129,4.1860)\dd \put(6.1604,4.3186)\dd
                       \put(6.4631,4.5745)\d  \put(6.7771,4.8161)\d
\put(7.1035,5.0401)\d  \put(7.4400,5.2406)\d  \put(7.7908,5.4145)\d
\put(8.1567,5.5547)\d  \put(8.5374,5.6520)\d  \put(8.9306,5.6951)\d
\put(9.3304,5.6716)\d  \put(9.7259,5.5714)\d  \put(10.1021,5.3921)\d
\put(10.4451,5.1418)\d \put(10.7470,4.8359)\d \put(11.0068,4.4897)\d
\put(11.2277,4.1154)\d \put(11.4139,3.7208)\d \put(11.5697,3.3108)\d
\put(11.6986,2.8879)\d \put(11.8033,2.4522)\d \put(11.8859,2.0011)\d
\put(11.9475,1.5247)\d \put(11.9888,.9764)\d
\end{picture}  \vspace{-.7cm}
\caption[{\frm Fig. 1.}]{{\frm Longitudinal plasma frequency
    versus coupling (schematically). The straight line corresponds to
    the leading order $\omega=m$. The curve inside the window is the
    subject of this paper. By the dots outside, the function
    $g \wu {1-g} $ is formally followed up to stimulate speculations. }}
\end{figure}    }
\def\fgrb{                      
\def\d{{$\!$.}}
\begin{figure}[t] \unitlength1cm \begin{picture}(15.2,9)
\put(2.7,6.5){1}  \put(0,7.5){\line(1,0){.5}}    
\put(2.5,7.5){\line(1,0){.5}}  \put(.5,7.5){\line(1,0){2}}
\put(1.5,7.5){\oval(2,1)[t]}   \put(1.5,7.5){\oval(2,1)[b]}
\put(6.7,6.5){2}                                 
\put(4,7.5){\line(1,0){.5}}  \put(6.5,7.5){\line(1,0){.5}}
\put(5,7.5){\circle{1.04}}   \put(6,7.5){\circle{1.04}}
\put(10.7,6.5){3}  \put(8,7.5){\line(1,0){.5}}   
\put(10.46,7.5){\line(1,0){.5}}  \put(9.5,7.5){\oval(2,1)[bl]}
\put(9.46,7.5){\oval(2,1)[r]}    \put(8.9,7.9){\circle{1.2}}
\put(15,6.5){4}                                
\put(12.04,7.5){\line(1,0){.44}}   \put(14.8,7.5){\line(1,0){.5}}
\put(13.9171,7.5)\d  \put(13.893,7.683)\d  \put(13.822,7.854)\d
\put(13.71,8)\d      \put(13.564,8.112)\d  \put(13.393,8.183)\d
\put(13.21,8.207)\d  \put(13.027,8.183)\d  \put(12.856,8.112)\d
\put(12.71,8)\d      \put(12.598,7.854)\d  \put(12.527,7.683)\d
\put(12.5,7.5)\d     \put(12.527,7.317)\d  \put(12.598,7.146)\d
\put(12.71,7)\d      \put(12.856,6.888)\d  \put(13.027,6.817)\d
\put(13.21,6.793)\d  \put(13.393,6.817)\d  \put(13.564,6.888)\d
\put(13.71,7)\d      \put(13.822,7.146)\d  \put(13.893,7.317)\d
\put(13.8,7.5){\oval(2,.8)[r]}
\put(2.7,3.5){5}  \put(0,4.5){\line(1,0){.48}}   
\put(2.52,4.5){\line(1,0){.48}}  \put(1.5,4){\line(0,1){1}}
\put(1,5)\d          \put(.809,4.962)\d    \put(.646,4.854)\d
\put(.538,4.691)\d   \put(.5,4.5)\d        \put(.538,4.309)\d
\put(.646,4.146)\d   \put(.809,4.038)\d    \put(1,4)\d  
\put(2,5)\d          \put(2.191,4.962)\d   \put(2.354,4.854)\d
\put(2.462,4.691)\d  \put(2.5,4.5)\d       \put(2.462,4.309)\d
\put(2.354,4.146)\d  \put(2.191,4.038)\d   \put(2,4)\d  
\put(1.2,5.01)\d \put(1.4,5.02)\d \put(1.6,5.02)\d \put(1.8,5.01)\d
\put(1.2,3.99)\d \put(1.4,3.98)\d \put(1.6,3.98)\d \put(1.8,3.99)\d
\put(6.7,3.5){6}  \put(4,4.5){\line(1,0){.498}}  
\put(6.5,4.5){\line(1,0){.5}}  \put(5.5,4){\line(0,1){1}}
\put(5.5,4.5){\oval(2,1)[t]}   \put(5.5,4.5){\oval(2,1)[b]}
\put(10.7,3.5){7}  \put(8,4.5){\line(1,0){.5}}   
\put(10.5,4.5){\line(1,0){.5}}  \put(9,3.99){\line(1,0){1}}
\put(9.5,5){\circle{1}}
\put(9,4.5){\oval(1,1)[l]}   \put(10,4.5){\oval(1,1)[r]}
\put(14.7,3.5){8}  \put(12,4.5){\line(1,0){.5}}  
\put(14.5,4.5){\line(1,0){.5}}  \put(13,3.99){\line(1,0){1}}
\put(13.5,5.5)\d      \put(13.309,5.462)\d  \put(13.146,5.354)\d
\put(13.038,5.191)\d  \put(13,5)\d          \put(13.038,4.809)\d
\put(13.146,4.646)\d  \put(13.309,4.538)\d  
\put(13.5,4.5)\d      \put(13.691,5.462)\d  \put(13.854,5.354)\d
\put(13.962,5.191)\d  \put(14,5)\d          \put(13.962,4.809)\d
\put(13.854,4.646)\d  \put(13.691,4.538)\d  
\put(13,4.5){\oval(1,1)[l]}      \put(14,4.5){\oval(1,1)[r]}
\put(2.7,0){9}   \put(0,1){\line(1,0){.48}}      
\put(2.52,1){\line(1,0){.48}}   \put(1.5,1.5){\oval(1,1)[t]}
\put(1,1.5)\d       \put(.809,1.462)\d   \put(.646,1.354)\d
\put(.538,1.191)\d  \put(.5,1)\d         \put(.538,.809)\d
\put(.646,.646)\d   \put(.809,.538)\d    \put(1,.5)\d   
\put(2,1.5)\d       \put(2.191,1.462)\d  \put(2.354,1.354)\d
\put(2.462,1.191)\d \put(2.5,1)\d        \put(2.462,.809)\d
\put(2.354,.646)\d  \put(2.191,.538)\d   \put(2,.5)\d   
\put(1.038,1.309)\d \put(1.146,1.146)\d  \put(1.309,1.038)\d
\put(1.5,1)\d       \put(1.962,1.309)\d  \put(1.854,1.146)\d
\put(1.691,1.038)\d  
\put(1.2,.49)\d \put(1.4,.48)\d \put(1.6,.48)\d \put(1.8,.49)\d
\put(6.6,0){10}  \put(4,1){\line(1,0){.498}}     
\put(6.5,1){\line(1,0){.5}}   \put(5.5,2){\circle{1}}
\put(5.5,1){\oval(2,1)[t]}    \put(5.5,1){\oval(2,1)[b]}
\put(9.6,0){11}  \put(8,1){\line(1,0){2}}        
\put(9,1.7){\oval(1.4,1.4)[b]}    \put(8.5,1.7){\oval(.4,.6)[tl]}
\put(9.5,1.7){\oval(.4,.6)[tr]}   \put(9,2){\circle{1}}
\put(12.4,0){12}                                 
\put(10.8,1){\line(1,0){2}}       \put(11.8,1.7){\oval(1.4,1.4)[b]}
\put(11.3,1.7){\oval(.4,.6)[tl]}  \put(12.3,1.7){\oval(.4,.6)[tr]}
\put(11.8,2.5)\d       \put(11.609,2.462)\d  \put(11.446,2.354)\d
\put(11.338,2.191)\d   \put(11.3,2)\d        \put(11.338,1.809)\d
\put(11.446,1.646)\d   \put(11.609,1.538)\d  
\put(11.8,1.5)\d       \put(11.991,2.462)\d  \put(12.154,2.354)\d
\put(12.262,2.191)\d   \put(12.3,2)\d        \put(12.262,1.809)\d
\put(12.154,1.646)\d   \put(11.991,1.538)\d  
\put(14.8,0){13}  \put(13.6,1){\line(1,0){1.6}}   
\put(14.4,1.66){\circle{1.4}}   \put(14.4,2.6){\circle{.7}}
\end{picture}
\caption[{\frm Fig. 2.}]{{\frm 2-loop diagrams.
    The dotted lines represent ghost propagators. Normal lines refer
    to hard and therefore bare gluons. The right-left mirror images
    of nos. 3 and 4 are included by doubling the corresponding symmetry
    factors. These are 1/6 (diagram 1), 1/4 (diagrams 2, 11, 13),
    1/2 (diagrams 6, 7, 10, 12), 1 (diagrams 3, 5, 8) and 2
    (diagrams 4, 9). }}
\end{figure}    }
\def\fgrc{                      
\def\d{{$\!$.}}
\begin{figure}[h,t]
\unitlength.8cm \begin{picture}(15.9,6.6)
\thicklines
\put(3,.5){\vector(1,0){12}}   \put(15.2,.45){$x$}
\put(9,.5){\vector(0,1){5.5}}  \put(8.9,6.3){$p$} \thinlines
\put(9,.5){\vector(1,1){5.5}}  \put(14.6,6.1){$a$}
\put(9,.5){\vector(-1,1){5.5}} \put(3.15,6.1){$b$}
\put(10,1.5){\line(-1,1){4.5}} \put(5.2,6){$^{a=1}$}
\put(11,2.5){\line(0,1){3.4}}  \put(10.7,6){$^{x=1}$}
\put(6.7,4.2){\line(1,1){.6}}   \put(7.2,4.77){$^{b=3}$}
\put(6.9,4.4){\line(-1,1){1.3}} \put(7.1,4.6){\line(-1,1){1.3}}
\put(7,.52)\d        \put(6.9940,.7)\d    \put(6.9760,.9)\d
\put(6.9462,1.1)\d   \put(6.9045,1.3)\d   \put(6.8513,1.5)\d
\put(6.7868,1.7)\d   \put(6.7111,1.9)\d   \put(6.6247,2.1)\d
\put(6.5279,2.3)\d   \put(6.4211,2.5)\d   \put(6.3047,2.7)\d
\put(6.1793,2.9)\d   \put(6.0454,3.1)\d   \put(5.9035,3.3)\d
\put(5.7541,3.5)\d   \put(5.5979,3.7)\d   \put(5.4355,3.9)\d
\put(5.2673,4.1)\d   \put(5.0941,4.3)\d   \put(4.9164,4.5)\d
\put(4.7347,4.7)\d   \put(4.5496,4.9)\d   \put(4.3615,5.1)\d
\put(9,.52)\d        \put(8.9940,.7)\d    \put(8.9760,.9)\d
\put(8.9462,1.1)\d   \put(8.9045,1.3)\d   \put(8.8513,1.5)\d
\put(8.7868,1.7)\d   \put(8.7111,1.9)\d   \put(8.6247,2.1)\d
\put(8.5279,2.3)\d   \put(8.4211,2.5)\d   \put(8.3047,2.7)\d
\put(8.1793,2.9)\d   \put(8.0454,3.1)\d   \put(7.9035,3.3)\d
\put(7.7541,3.5)\d   \put(7.5979,3.7)\d   \put(7.4355,3.9)\d
\put(7.2673,4.1)\d   \put(7.0941,4.3)\d   \put(6.9164,4.5)\d
\put(6.7347,4.7)\d   \put(6.5496,4.9)\d   \put(6.3615,5.1)\d
\put(11,.52)\d       \put(11.0060,.7)\d   \put(11.0240,.9)\d
\put(11.0538,1.1)\d  \put(11.0955,1.3)\d  \put(11.1487,1.5)\d
\put(11.2132,1.7)\d  \put(11.2889,1.9)\d  \put(11.3753,2.1)\d
\put(11.4721,2.3)\d  \put(11.5789,2.5)\d  \put(11.6953,2.7)\d
\put(11.8207,2.9)\d  \put(11.9546,3.1)\d  \put(12.0965,3.3)\d
\put(12.2459,3.5)\d  \put(12.4021,3.7)\d  \put(12.5646,3.9)\d
\put(12.7327,4.1)\d  \put(12.9059,4.3)\d  \put(13.0836,4.5)\d
\put(13.2653,4.7)\d  \put(13.4504,4.9)\d  \put(13.6385,5.1)\d
\end{picture}  \vspace{-.7cm}
\caption[{\frm Fig. 3.}]{{\frm Integration area in the numerical
evaluation of $\eta_4$. On the dotted lines there are poles of
the spectral density (outside) or of the propagator. The stripe
on the line $a=1$ had a width of $\Delta a =.2$. }}
\end{figure}    }
\documentstyle[12pt,twoside]{article}

\let\thq=\theequation        \font\frm=cmr10
\def\mn{ _{\mu \nu} }        \def\omn{ ^{\mu \nu} }
\def\be{ \begin{equation} }  \def\bea{ \begin{eqnarray} }
\def\ee{ \end{equation} }    \def\eea{ \end{eqnarray} }
\def\P{ {\mit\Pi} }          \def\nonu{ \nonumber }
\def\dis{ \displaystyle }    \def\Tr{ \, {\rm Tr} \, }
\def\parag#1{ \vspace{1.5cm} \hspace{.08cm} \parbox{15cm}{{\bf #1}
       \vspace{1.1cm} } \hfill \vphantom{a} \nopagebreak \indent }
\def\wu#1{\sqrt{{#1} \,}^{ \hbox
   to0.2pt{\hss$ \vrule height 2pt width 0.6pt depth 0pt $} \;\! } }
\begin{document}  
\begin{titlepage}
\begin{flushright}   DESY 93-080 \, , \, ITP-UH 8/93 \\
                     June 1993 \, , \, Revised September 1993
\end{flushright}    \vfill  \vskip 1.5cm
\begin{center}
{\Large \bf         Gluon Plasma Frequency -- } \\ \medskip \smallskip
{\Large \bf         the Next-to-Leading Order Term } \\
\vskip 1.3cm \vfill
\vfill    {\large   Hermann Schulz}\\
\bigskip  {\sl      Institut f\"ur Theoretische Physik,
                                         Universit\"at Hannover \\
                    Appelstra\ss e 2, D-30167 Hannover, Germany \\  }
\vskip 2.5cm \vfill                    {\large    ABSTRACT}
\end{center}  \begin{quotation}  \ \ \
\end{quotation}  \end{titlepage}
   The longitudinal-electric oscillations of the hot gluon system
   are studied beyond the well known leading order term at high
   temperature $T$ and small coupling $g$. The coefficient
   $\eta$ in $\omega^2 = m^2 \, (1+ \eta \, g \wu N \, )$ is
   calculated, where \hbox{$\omega \equiv \omega (\vc q =0)$} is
   the long-wavelength limit of the frequency spectrum, $N$ the
   number of colours and $m^2=g^2 N T^2/9$. In the course of this,
   for the real part of the gluon self-energy, the Braaten-Pisarski
   resummation programme is found to work well in all details. The
   coefficient $\eta$ is explicitly seen to be gauge independent
   within the class of covariant gauges. Infrared singularities
   cancel as well as collinear singularities in the two-loop
   diagrams with both inner momenta hard. However, as it turns out,
   none of these two-loop contributions reaches the relative order
   $O(g)$ under study. The minus sign in our numerical result
   $\; \eta = -.18 \; $ is in accord with the intuitive picture
   that the studied mode might soften with increasing coupling
   (lower temperature) until a phase transition is reached at
   zero-frequency. The minus sign thus exhibits the 'glue' effect
   for the first time in a dynamical quantity of hot QCD.
%
%
\let\dq=\thq \renewcommand{\theequation}{1.\dq}
\setcounter{equation}{0}

\parag {1. \ Introduction }
Mostly, our understanding of a complex physical problem profits from
its known solution at the end of some parameter axis. In the case of
QCD the large-N limit so far failed in 3+1 dimensions: the 'master
field' is not known \cite{cole}. But for QCD in contact with a
thermal bath there is indeed such a parameter and is called
temperature. It appears that during the last years the essential
problems with the high temperature limit of QCD have been overcome.
The coupling $g$ is weak there, perturbation theory is applicable,
and the limiting form of several quantities (as e.g. the two-gluon
Greens function at ingoing momentum $\sim gT$) can be written down
explicitly in this limit. The above prospect is one of the reasons
for the current high interest in hot QCD. The more immediate reasons
are the relevance to heavy ion collisions and to the early universe.

Several difficulties specific to thermal gauge theory \cite{hotgauge}
long hampered the above understanding of temperature as a useful tool.
Especially for the damping rate of the gluon plasma oscillations
various numbers (mostly negative, hence unphysical) were produced,
roughly one for each gauge used \cite{K+K,GPY,samm}. In contrast,
the gauge independence of plasma parameters was demonstrated
nonperturbatively \cite{kokure}, ranking them among the measurable
physical quantities. It turned out that resummation is inevitable.
However, even if gauge independence is restored on-shell by
resummation \cite{KRS}, the latter may be incomplete.

The breakthrough came with a few papers of Pisarski
\cite{pispect,pi89} and Braaten and Pisarski \cite{BP,BPward} around
1990. In their basic paper (which is \cite{BP} and henceforth
referred to as BP) it was shown that, in order to obtain soft
amplitudes consistently, hard thermal loops must be added to the
tree-level vertices and summed up in the gluon propagator. A
momentum $\sim T$ is 'hard', but if $\sim gT$ it is 'soft'
($ 0 < g <\!\! < 1 $). Gauge independence of this setup was proved
\cite{BP,FT} and the (gauge independent and positive) gluon damping
rate was obtained \cite{6.6}. The development culminated in giving
this new ''true zeroth order'' the form of a Lagrangian
\cite{eff,shortcut}, which generates the leading terms of all soft
amplitudes and can be even rewritten in a mainfestly gauge-invariant
form \cite{BPeff}. Applications cover the soft dilepton production
\cite{dilep}, quark damping \cite{qdamp}, screening \cite{BY}, energy
loss \cite{BTo}, kinetic equations \cite{blaiz} or even star matter
\cite{astro}. There are current questions concerning the existence
of a 'magnetic mass' \cite{GPY,AKR}, the measurability of the
damping \cite{smil} and the regulator, which prevents revived gauge
dependence of the damping. But through the present work we were not
forced into the former problems. With regard to the latter, all
gauge dependences are cancelled algebraically on the plasmon
mass-shell. We assume that potential mass-shell singularities
\cite{baier} are regularized in the manner of ref. \cite{tobks}.

The system considered in this paper consists of only gluons in
thermal equilibrium (no quarks). We concentrate on the real part of
the frequency of the plasmon mode and take the first step beyond
'zeroth order'. There are three possible origins of contributions
to the relative order $O(g)$ ('relative' means up to the prefactor
$m^2 = g^2 T^2 N/9 $). These origins form the section headings of
\S\S\, 3, 4 and 5. Their classification is due to BP. Thus, the best
introduction to the present paper is the subsection 4.3 in BP. We
shall not summarize this paragraph here. The predictions of BP
concern the possible maximum contribution to each subset. Explicit
calculation may well give something below $O(g)$ (a) for kinematical
reasons, (b) by 'accidental' cancellation of prefactors and (c) by
compensation among ranges of the integrals. BP give an example for
case (a) in discussing the imaginary part of hard one- and two-loop
diagrams. Sections 3 and 4 give examples for case (c).

As we restrict ourselves to the long-wavelength limit $\vc q
\rightarrow 0$ there remains one single number to be calculated. This
number is the prefactor $\eta$ in (\ref{etom}) below. It is related
to the real part of that quantity $\gamma = \gamma_r + i \gamma_i$
whose imaginary part is the damping rate:
\be  \label{etom}
  \omega^2 = m^2 \left( 1+\eta g \wu N \right) \equiv m^2
  - 2m\gamma_r \quad ,\quad \gamma = a \, {g^2 N T \over 24 \pi }
  \quad \Rightarrow \quad a_r = - 4\pi \eta \;\; . \quad
\ee
Hence, our result $\eta = -.18$ (already disclosed in the abstract)
means $a_r = 2.3$. The real part is thus somewhat smaller than
the imaginary part $a_i = 6.635$ \cite{6.6}. The second digit
in $-.18$ is not quite certain. The work leading to this number
is laborious. Our motivations were:
\begin{itemize}
\item[(i)\,\,]
   Is QCD physically simple? If so, our expectation on the behaviour
   of e.g. the gluon system should be confirmed immediately in the
   first term of the 'true' perturbation expansion. We expect that,
   with increasing coupling, the 'glue' reduces the frequency of
   the plasmon mode below its zeroth-order value $m$. Moreover, this
   frequency could play the role of an indicator, reaching zero at
   the onset of glue ball formation. In figure 1, an increasing
   coupling might be associated with decreasing temperature
   \cite{enqv}. Remember that often (especially in asymptotic series)
   the first term of a perturbation expansion gives qualitatively the
   full answer.
\item[(ii)\,\,]
   The high-temperature limit as a perturbative starting point needs
   examples. We should like to give one more. The first example was
   given already in 1979 as Kapusta \cite{kapudruck} calculated the
   pressure $p\sim T^4 \left( 1 - 5 g^2 N / 16 \pi ^2 \right) $. Note
   the minus sign.
\item[(iii)\,\,]
   BP at work. While filling \S 4.3 of BP with detail, we will test
   the resummation programme independently. The test concerns the
   separate gauge independent sets within $O(g)$, but also the
   absence of infrared singularities, UV-convergence and (last not
   least) the physics, which here is in the minus sign in question.
\item[(iv)\,\,]
   Working with the BP resummation we shall reformulate
   it in our Minkowski notation. This is a matter of language only,
   we do not claim for preferences. BP use 'English', say, and here
   is the translation into 'Dutch' \cite{LW}.
\end{itemize}
      \fgra    

The paper is organized as follows. It starts with the details of the
'zeroth approximation' (section 2). In the next three sections the
$O(g)$ contributions are calculated. We follow the BP classification
in reverse order. The two-loop diagrams (section 3) could be
suspected to be outside of the realm of feasability. But they are
not. Both, the two loop diagrams and the 1-loop hard diagrams
(section 4) do not (yet) contribute to $O(g)$. The main part is
section 5 on the one-loop soft diagrams. The formal result of the
soft analysis is summarized in section 6. Here we go until to the end
of the analytical treatment. In section 7 the figure 3 gives a rough
view into the numerical procedure.

\newpage
%
%
\let\dq=\thq \renewcommand{\theequation}{2.\dq}
\setcounter{equation}{0}

\parag {2. \ The frequency of the plasmon mode }
In this short section we specify the subject and introduce notations.
For simplicity, we allow for only gluons, activated thermally and of
$N^2-1$ kinds. To get rid of quarks in a physical manner, they would
have to be given masses much larger than the temperature. Then, the
Lagrangian reads
\be  \label{lagrangian}
  {\cal L} = - {1 \over 4} \, F\mn^{\,\enskip a} F^{\mu \nu \; a} -
  {1 \over 2 \alpha} \,\left( \partial^\mu \! A_\mu^a \right)^2
  \;\; + \;\; \mbox{ghost term} \quad .
\ee
We use the Matsubara contour and Minkowski metric \hbox{$+ - - -$}
\cite{LW}. Hence a four vector reads
$P = ( \, i\omega_n \, , \, \vc p \, )$ with $\omega_n = 2\pi n T$,
$P^2=(i\omega_n)^2 -p^2$. Let $Q$ be the argument of the polarization
function (its 'outer momentum'). We shall keep writing $Q_0$ even if
it is already continued into the complex plane. By now the term '$Q$
soft' applies to $Q_0$ as well.

Although the terms 'gluon self energy' and 'polarization function'
have identical meaning, we prefer the latter to emphasize the view
of a medium having dielectric properties. The longitudinal plasmon
mode (which lives on degrees of freedom not activated at zero
temperature \cite{landreb}) is detected as a zero of the dielectric
constant or, equivalently, as a pole of the longitudinal part (index
$\ell$) of the gluon propagator. To 'zeroth order', i.e. when dressed
with hard thermal loops, and within covariant gauges this propagator
reads
\bea   \label{gmn}
  G\mn (P) &=& A\mn (P) \Delta_t (P) + B\mn (P) \Delta_\ell (P)
             + D\mn (P) \Delta_\alpha  \\
       \label{deltas}
 \mbox{where} & & \Delta_\alpha = \alpha \Delta_0 \;\; , \;\;
  \Delta_0 = {1 \over P^2} \;\; , \;\; \Delta_{t,\, \ell} =
  { 1 \over P^2 - \P _{t,\, \ell } (P) }  \;\; .
\eea
We emphasize that the above object (at soft momentum $P$) is much
more than a certain perturbative outcome with uncertain meaning. It
is, as BP have demonstrated, the exact asymptotically leading term of
the propagator in the limit of high temperature ($gT <\!\! < T$). The
ghost propagator is $\Delta_0$ and remains undressed to zeroth order.
The Lorentz-matrices in (\ref{gmn}) belong to the matrix-basis
\cite{GPY,K+K,LW,ich}:
\bea  \label{A-D}   & & \hspace{-1cm}
  A= g-B-D \;\; , \;\; B= { V \circ V \over  V^2 } \;\; , \;\;
  C= { Q \circ V + V \circ Q  \over \wu 2 Q^2 q } \;\; , \;\;
  D= { Q \circ Q \over  Q^2 } \qquad \\  \label{V} & & \hspace{-1cm}
  \mbox{with}  \quad V= Q^2 U - (U \cdot Q) \, Q = ( - q^2 \, ,
  \, - Q_0 \vc q \, ) \;\; , \qquad
\eea
where $U=(1 \, , \vc 0 \, )$ is the four-velocity of the thermal
bath at rest. The form (\ref{gmn}) derives from $G=G^0 + G^0 \P G$
by using (\ref{A-D}) and the polarization function $\P\omn$ at
one-loop order (leading term as given in (\ref{piyy}) below).

At this point it is clear how the position of the pole in the
$B$-term of the propagator is obtained to any desired higher order
in the coupling: consider the corresponding 1PI diagrams, which form
$\P\omn$, but formulate them with dressed lines (\ref{gmn}) and
dressed vertices (see below) and consider counter terms, see \S 4.3
of BP, which here, however, are not yet needed. Also $n$-vertices
with $n>4$ do not yet occur. Once $\P\omn$ is obtained that way,
one forms $\P_\ell = \Tr B \P$. If one is interested in the limit
$\vc q \rightarrow 0$ only, one obtains $\omega \equiv \omega
(\vc q = 0)$ as $\, \omega = \Re e \, \Omega \,$ by solving
\be   \label{bigom}
  \Omega^2 = \P_\ell \, ( \Omega , \vc q = 0 )
\ee
for the complex number $\Omega$. At next-to-leading order in $g$
this equation reduces to $\omega^2 = \Re e \; \P_\ell (m,0)$
with $m= g\wu N T/3$. On dimensional grounds $\omega$ is $m$ times
some function $f$ of only $g$. Thus, the only explicit temperature
dependence of $\omega$ is the trivial one in the prefactor $m$; the
other $T$-dependence is implicit in the running of the coupling $g$.
At small $g$ the function $f(g)$ need not be a pure power series in
$g$. Possibly the asymptotics of $f$ looks as follows,
\be  \label{logg}
 \omega^2 = m^2 \left( \, 1 + \eta \, g \wu N + \overline{\eta}
 \, g^2 \ln (g) + \overline{\overline{\eta}} \, g^2 + \ldots
 \;\right)  \;\; ,
\ee
because such logarithmic terms will appear in section 4. Since
$\omega$ is a measurable quantity, {\sl each} term of its asymptotics
must be gauge independent. If not, the calculation is wrong.

We continue listing further details on the 'zeroth order'. The
leading term \cite{BPward,shortcut} of $\P\omn$ is
\bea \label{piyy}
  \P\omn (Q) &=& 3 m^2 \left( \, U^\mu U^\nu - < \, (U\cdot Q )
           \, Y^\mu Y^\nu \, / \, (Y \cdot Q ) \, > \,\right) \\
  \label{pino}
         &=& {3 \over 2} m^2 g\omn + 4 g^2 N
             \sum { K^\mu K^\nu \over K^2 (K-Q)^2 } \;\; ,
\eea
where $Y \equiv (1,\vc e )$, $Y^2=0$. $< \ldots >$ is the
average over the directions of the unit vector $\vc e $, and the
blank summation symbol means
\be \label{sid}
    \sum_K \; \equiv \; \int_K^3 \sum_{K_0} =
    \int_K^3 \; T \sum_n = \left( {1 \over 2 \pi}
    \right) ^3 \int \! d^3 k \;\, T \sum_n  \;\; .
\ee
We write $n(k)= 1/(e^{\beta k} -1)$ for the Bose function and
$q^\ast = T \wu g $ for the threshold between hard and soft momenta
\cite{BY}. Since the functions $\P_t = \Tr A \P /2 $ and
$\P_\ell =\Tr B\P $ in (\ref{deltas}) are related by
\be \label{l+t}
   \P _\ell (Q) + 2 \P _t (Q) = 3 m^2
\ee
(use (\ref{piyy}) together with (\ref{A-D})), we have to record only
\be \label{piell}
 \P _\ell (Q) = 4g^2 N \sum {1 \over K^2 (K-Q)^2 } \left[
             k^2 - { (\vc k \vc q )^2 \over q^2 } \right] \;\; .
\ee
For this sum evaluated see Appendix B. There also the spectral
densities of the propagators $\Delta_t$ and $\Delta_\ell$ are detailed.
For the definition of spectral densities and the general spectral
representation see (\ref{5spect}) below. Often differences of two
propagators do occur:
\be \label{deldiff}
  \Delta_{\ell \, t} \equiv \Delta_\ell - \Delta_t \quad , \quad
  \Delta_{\alpha \, \ell} \equiv \alpha \Delta_0 - \Delta_\ell\;\; .
\ee

Last not least, if the outer momenta are soft (as in section 5), the
3- and 4-vertices \cite{BP} are to be dressed by one hard thermal
loop each. After the colour sums are done, the remaining parts of
the vertices read as follows (cf. e.g. (3.2) and (3.28) in BP, the
different sign is due to notation):
\bea \label{gam3}
 \Gamma^{123} &=& (Q_1 \vert Q_2 \vert Q_3 )^{123} + \delta
 \Gamma^{123} \;\; , \\
 \mbox{where} & & \delta \Gamma^{123} = - 8 g^2 N \sum_K { K^1
    K^2 K^3 \over K^2 (K+Q_1 )^2 (K+Q_1 + Q_2 )^2 }  \nonu \\
\label{qqq} \mbox{and} & & (Q_1 \vert Q_2 \vert Q_3 )^{123}
    = (Q_1 - Q_2 )^3 g^{12} + \;\; \mbox{cyclic}  \;\; ;
\eea
\bea \label{gam4}
 \Gamma^{1234} &=& g^{14} g^{23} + g^{13} g^{24} - 2 g^{12} g^{34}
                + \delta \Gamma^{1234} \;\; , \\
 \mbox{where} & & \delta \Gamma^{1234} = 16 g^2 N \sum_K K^1
     K^2 K^3 K^4 \left( {2 \over N_{1234}} + {2 \over N_{2134}}
     + {1 \over N_{4231}} \right) \nonu \\
 \mbox{and} & & N_{1234} = K^2 (K+Q_1 )^2 (K+Q_1 + Q_2 )^2 (K-Q_4)^2
     \;\; .\nonu
\eea
In both expressions, (\ref{gam3}) and (\ref{gam4}), the sum of the
$Q_i$ must vanish. The dressed-vertex Ward identities, cf. (3.31) and
(3.33) of BP, are
\bea \label{ward3}
 (Q_3)_3 \delta \Gamma^{123} &=& \P ^{12} (Q_1) - \P ^{12} (Q_2)  \\
 \label{ward4}
 (Q_4)_4 \delta \Gamma^{1234} &=& -\delta \Gamma^{123} (Q_1 + Q_4 ,
    Q_2 , Q_3 ) + \delta \Gamma^{123} (Q_1 , Q_2 + Q_4 , Q_3 ) \;\; .
\eea
This completes our listing of known 'zeroth order' results as we
shall need them.

The plasmon mode is a 'longitudinal-electric' wave. To appreciate
this term (used in the abstract) note that $V^\mu A_\mu^a (Q) = -i
\vc q \cdot \vc E ^a (Q) \equiv -i q E_{\rm long}^a (Q) $, where
$\vc E ^a (Q)$ is the Fourier transform of $- \partial_0 \vc A^a (x)
- \nabla A_0^a (x)$. Herewith the $B$-term of the full gluon
propagator may be written as
\be
     {1 \over Q^2} < E^a_{\rm long} (Q) \; E^b_{\rm long} (-Q) >
     \;\; = \; { \delta^{a b} \over Q^2 - \overline{\P}_\ell (Q) }
\ee
with $\overline{\P}_\ell $ the exact longitudinal polarization
function and $< \ldots >$ the thermal average. The strongly
correlated fields near the pole are indeed longitudinal electric ones.
%
%
\let\dq=\thq \renewcommand{\theequation}{3.\dq}
\setcounter{equation}{0}

\parag {3. \ Two-loop diagrams with hard inner momenta} In this
section the complete set of 2-loop diagrams is analysed with respect
to its possible $g^3$-contribution to the real part of the
polarisation function $\P\omn (Q)$. Here, and only here, we shall
restrict ourselves to Feynman gauge ($\alpha=1$). The outer momentum
reads $Q=(Q_0,\vc q )$, and the limit $q \rightarrow 0$ is taken as
early as possible. We emphasize these restrictions, although we do
not expect them to be crucial for the somewhat unexpected results,
namely, that the 2-loop contributions turn out to remain below the
relative order $O(g)$. Because of the latter, readers who are only
interested in the relevant terms might skip this section right now.

The set of 2-loop diagrams is shown in figure 2. They are numbered
from $i=$1 to $i=$13. Correspondingly, there is an $i$-th
contribution to $\P $, and each has an individual numerator $n \omn$
and denominator $d$ under a double sum over the hard inner momenta
$P$ and $K$:
\be \label{H}
  \P \omn = {1\over 2} g^4 N^2 \sum \sum { n \omn \over d } \;\;\; ,
   \quad \P _\ell = \Tr (B \P ) \equiv {1 \over 2} g^4 N^2 H
   \;\; , \;\;\; H = \sum \sum {n_\ell \over d} \;\; .
\ee To be specific, the denominators $d$ are
\def\ii#1{ \,\; (\, i = #1\, ) }
\bea \label{ds}
   (K-Q)^2 (P-K)^2 P^2      \ii {1}     & , &
   K^4 (K-Q)^2 (P-K)^2 P^2  \ii {7,8,9}  \nonu \\
   K^2 (K-Q)^2 P^2 (P-Q)^2  \ii {2}     & , &
   K^4 (K-Q)^2 P^2          \ii {10}     \nonu \\
   K^2 (K-Q)^2 (P-K)^2 P^2  \ii {3}     & , &
   K^4 (P-K)^2 P^2          \ii {11,12}  \nonu \\
   K^2 (K-Q)^2 (P-K)^2 P^2 (P-Q)^2 & & \hskip -.7cm \ii {4,5,6}
\;\;\; , \;\;\;  K^4 P^2    \ii {13}  \;\; .
\eea
In three denominators (nos. 2, 10 and 13) the factor $(P-K)^2$ is
absent. In these cases the two sums are easily decoupled. The
symmetry factors are given in the figure caption. They are included
in $n \omn$ and $n_\ell \equiv \Tr (B n)\,$.

To exhibit the typical steps in treating any of the more complicated
diagrams we shall work out one example in detail. The results for
the 12 others will then be listed only. Consider the loop with an
inserted ghost loop: number 8. The symmetry factor is 1. The
structure constants at the ghost vertices combine via $f^{\circ
\bullet a} f^{\circ \bullet b} = N \delta^{a b}$, and the Kronecker
helps to treat the remaining two $f$'s in the same manner. Using the
notation (\ref{qqq}):
\be \label{n1}
  n \omn = - 2 (Q \left| \, K-Q \, \right| -K)^{\mu \lambda \rho}
    (P-K)_\rho {(Q \left | \, K-Q \, \right| -K)^\nu_{\;\,
    \lambda} }^{\; \sigma}  P_\sigma \;\; .
\ee
At this point it is convenient to leave the algebra to a little
REDUCE program. Nevertheless, all calculations have been checked
by hand.
      \fgrb    

The denominator $d=K^4 (K-Q)^2 (P-K)^2 P^2$ does not change under
the substitution $P \rightarrow K-P$. Thus, the numerator
$n \omn$ may be replaced by its symmetric part under this
transformation. Once symmetric, it can be expressed by invariants
$\cal I$ or by pairs of 'odd-invariants' $\cal O$ (which change
sign under $P\rightarrow K-P$)$\,$. Such invariants are
\bea \label{inv1}
 {\cal I} = 2P^2-2PK \;\; , \;\; {\cal I} \omn
          = 2 P^\mu P^\nu - P^\mu K^\nu
            - K^\mu P^\nu \;\; , \hskip 3.5cm  \nonu \\
 {\cal O}_K = 2 PK - K^2 \;\; , \;\; {\cal O}_Q= 2 PQ-KQ \;\; , \;\;
 {\cal O} \omn = P^\mu K^\nu + K^\mu P^\nu - K^\mu K^\nu \;\; .
\eea
The result for $n \omn $ then reads
\bea \label{n2}
n \omn & = & - {1 \over 2} g \omn ( {\cal O}_K - 2 {\cal O}_Q )^2 +
           {1 \over 2} g \omn \left( K^2-2KQ \right) ^2 -
           K^\mu K^\nu \left( 3 K^2 + 4 {\cal I} \right) \nonu \\
& & {} -{\cal I} \omn (K+Q)^2 + 3 {\cal O} \omn {\cal O}_K \;\; +\;\;
   {\rm terms \; containing} \; Q^\mu \; {\rm or/and} \; Q^\nu \;\; .
\eea
The terms not made explicit in (\ref{n2}) vanish due to $V \cdot Q=0$
under the $\Tr B \ldots $ operation. Conveniently, when taking this
trace (with ${\cal I} \omn $, say) we also exploit
$\vc q \! \rightarrow \! 0$, which amounts to $B \rightarrow - \,
(0,\vc q) \circ (0,\vc q ) / q^2 $. The two angular integrations in
$\sum \sum$ now permit the replacement ${\cal I} \omn \rightarrow
- {2 \over 3} \left( p^2 - \vc p \vc k \right) $. We obtain:
\bea \label{nb1}
n_\ell & = & - {1 \over 2} {\cal O}_K^2 + 2 {\cal O}_K {\cal O}_Q
        - 2 {\cal O}_Q^2 + {1 \over 2 } \left( K^2-2KQ \right) ^2
        + k^2 K^2   \nonu \\
 & & {} + {4 \over 3} k^2 {\cal I}
        + {2 \over 3} \left( p^2 - \vc p \vc k \right) (K+Q)^2
        - \left( 2 \vc p \vc k - k^2 \right) {\cal O}_k \;\; .
\eea
Note, that here and anywhere in the following $\vc q = \vc 0 $ and
hence $Q = (Q_0, \vc 0 )$. Next we try and rewrite $n_\ell$ as a
linear combination of factors which occur in the denominator:
\bea \label{inv2}
  {\cal I} & = & P^2+(P-K)^2-K^2 \;\; \rightarrow \;\; 2(P-K)^2 -K^2
   \nonu \\    {\cal O}_K & = & P^2 - (P-K)^2 \;\; \rightarrow \;\;
  - 2(P-K)^2 \nonu \\
  - {1 \over 2} {\cal O}_K ^2 & \rightarrow & (P-K)^2 (2PK-K^2)
                 \;\; \rightarrow \;\; - (P-K)^2 K^2      \nonu \\
 - 2 {\cal O}_Q^2 & = & -2 Q_0^2 (2P_0-K_0)^2 \;\; \rightarrow \;\;
   - 2 Q^2 \left( 4 (P-K)^2 -K^2 +4 p^2 - k^2 \right) \;\; .
\eea
The rightarrows in (\ref{inv2}) indicate allowed substitutions in
(\ref{nb1}). The symmetry is now abandoned in favour of cancellations.
Note that if a factor $(P-K)^2$ is cancelled with that in the
denominator there is another symmetry, namely $P \rightarrow -P$,
which allows for the last step in the third line of (\ref{inv2}). If
$(K-Q)^2$ is cancelled, the new symmetry is $K, P \rightarrow -K, -P$.
Through such steps we arrive at                \def\hh{\hspace{-1cm}}
\bea \label{nb2} \hspace{-.5cm}
 n_\ell &=& (P-K)^2 \left( K^2 -2 (K-Q)^2 + {2 \over 3} k^2 \right)
         + (K-Q)^2 \left( {1 \over 2} K^2 + {1 \over 3 } k^2
         - {2 \over 3} p^2 \right)   \nonu \\
 & & {} \hh + K^2 \left( {4 \over 3} p^2 - k^2 \right)
     + \, n_1 \, + \, n_2 \, + Q^2 \left( 2K^2 - {1 \over 2} (K-Q)^2
     + {1 \over 2} Q^2 - 6 (P-K)^2 \right)  \\
 & & \hh \; {\rm with} \quad n_1 = {4 \over 3} Q^2 k^2 \;\;\;
        {\rm and}  \quad n_2 = - {20 \over 3} Q^2 p^2 \;\; . \nonu
\eea
This is not the last version. So far we have done nothing towards the
fact that both inner momenta may be taken hard. Clearly, the last
lengthy term in (\ref{nb2}) is two $g$-orders smaller than e.g. the
first one. Hence, we neglect it. It is tempting to do so with the
term $n_1$ as well. This, however, is not allowed:
\bea \label{qk}
 4 Q^2 k^2 & = & 4 (KQ)^2 - 4 Q^2 K^2 \nonu \\
           & = & K^4 - K^2 (K-Q)^2 - 2 Q^2 K^2 + Q^4 - Q^2 (K-Q)^2
                 - (K-Q)^2 2 K Q    \nonu \\
     & \approx & K^4 - K^2 (K-Q)^2  \;\; ,
\eea
where the last term in the second line was omitted due to
$K,P \rightarrow -K,-P$. In the case of $n_2$ this last step does not
work. In fact, $n_2$ remains of the relative order $g^2$ and is to be
neglected. The hard-hard result for number 8 is now obtained:
\be \label{h8}
  H_8 = {4 \over 3} Z_2 - {2 \over 3} Z_2^\prime - Z_1 + {1 \over 3}
        Z_1^\prime + {1 \over 3 } Z_0 + {1 \over 6} Z_0^\prime
     + I_0 \left( {2 \over 3} J_1 + J_0 - 2 J_0^\prime \right) \;\; ,
\ee
where the objects $Z$, $I$, $J$ are sums out of the following
collection:
\bea \label{zs} \hspace{-.5cm}
      Z_0 = \sum \sum { 1 \over (K-Q)^2 (P-K)^2 P^2 }
 & , & Z_{1,\, 2} = \sum \sum { ( \, \vc k ^2 \, , \; \vc p ^2 \, )
            \over K^2 (K-Q)^2 (P-K)^2 P^2 } \;\; ,\nonu \\
      Z_0^\prime = \sum \sum { 1 \over K^2 (P-K)^2 P^2 }
 & , & Z_{1,\, 2}^\prime = \sum \sum { ( \, \vc k ^2 \, , \;
        \vc p ^2 \, ) \over K^4 (P-K)^2 P^2 } \;\; ,\nonu \\
       I_0 = \sum {1 \over K^2 } = - {T^2 \over 12 }
 & , & I_1 = \sum {\vc k ^2 \over K^2 (K-Q)^2 } = {T^2 \over 24 }
             \;\; , \nonu \\
       J_0 = \sum { 1 \over K^2 (K-Q)^2 }
 & , & J_1 = \sum { \vc k ^2 \over K^4 (K-Q)^2 } \;\; , \;\;\;
             J_0^\prime = \sum {1 \over K^4}  \;\; .
\eea

Something enervating happened in the last steps leading to (\ref{h8}).
Certain terms with a prefactor $Q^2$ were neglected but others not.
Consider again the term $n_1$ in (\ref{nb2}). It had the effect of
adding a term $(Z_0 -Z_0^\prime )$ to $H_8$ (moreover, $Z_0^\prime =0$,
see (\ref{z12}) below). With regard to (\ref{zs}) the two sums only
differ by the kind of pole prescription. Furthermore, if we replace
$n_\ell$ in (\ref{H}) by $n_1$, such an expression could well be among
the 3-loop contributions (read $Q^2$ as $\omega^2 \sim g^2 T^2$ and
the $T^2$ as e.g. $I_0\,$, i.e. as a loop that factorizes off). To
summarize, we learn that the correct definition of 2-loops requires
3-loops. Here we do what we can and evaluate the 2-loop terms as
they stand. Fortunately, as it will turn out shortly, the 2-loop
terms remain below the relevant order $O(g)$.

With the above mentioned reservation in mind we return to the full
set of all 13 diagrams and list the results:
\def\komm{ \quad , \quad }  \def\h{\hspace{-.7cm}}
\bea
\h H_1 & = & 9 Z_0 \komm H_2 \; = \; 0  \komm
   H_3 \; = \; 15 Z_1 -{27 \over 2} Z_0 + {9 \over 2} Z_0^\prime
            \;\; , \nonu\\
\h H_4 & = & {2 \over 3} Z_2 - {1 \over 3} Z_1 - {1 \over 6} Z_0
      + {1 \over 6} Z_0^\prime - {2 \over 3} I_1 J_0 \komm
   H_5 \; = \; {1 \over 6} Z_1 + {1 \over 12} Z_0
            - {1 \over 12} Z_0^\prime    \;\; ,\nonu \\
\h H_6 & = & 6 Z_2 - {41 \over 6} Z_1 + {43 \over 12} Z_0
         - {19 \over 12} Z_0^\prime - 6 I_1 J_0 \;\; ,\nonu \\
\h H_7 & = & - {20 \over 3} Z_2 + {10 \over 3} Z_2^\prime
       - {25 \over 3} Z_1 - {7 \over 3} Z_1^\prime
       + {8 \over 3} Z_0 + {5 \over 6} Z_0^\prime
       - 10 I_0 J_1 -3 I_0 J_0 + 12 I_0 J_0^\prime \;\; ,\nonu \\
\h H_8 & \! {\rm see }\! & (\ref{h8}) \komm
   H_9 \; = \; {2 \over 3} Z_1     \;\; ,\nonu \\
\h H_{10} & = & 20 I_0 J_1 -6 I_0 J_0 -6 I_0 J_0^\prime \komm
   H_{11} \; = \; - {10 \over 3} Z_2^\prime + {7 \over 3} Z_1^\prime
            - 5 Z_0^\prime - 16 I_0 J_0^\prime  \;\; ,\nonu \\
\h H_{12} & = & {2 \over 3} Z_2^\prime - {1 \over 3} Z_1^\prime
          - Z_0^\prime + 2 I_0 J_0^\prime  \komm
   H_{13} \; = \; 18 I_0 J_0^\prime \;\; .
\eea
In the first line, the result for $H_2$ actually was
$-27 Q^2 J_0^2/4 \,$, which however had to be neglected in the
hard-hard sense. To deal with all contributions, the collection
(\ref{zs}) was sufficient. But one of these sums diverges, namely
$Z_2^\prime$ (see below). Appeasingly enough, the $Z_2^\prime$-terms
cancel each other when adding $H_7$ to $H_{11}$, or $H_8$ to
$H_{12}$. Summing up the 13 contributions $H_i$ one obtains
\be \label{hsum} \hspace{-.4cm}
   \P _\ell ^{\; \mbox{\frm 2-loop hh}} = g^4 N^2 \left(
   {2 \over 3} Z_2 - {1 \over 3} Z_1 + Z_0 - Z_0^\prime
   - 4 I_0 (J_0 -J_0^\prime )
   + {16 \over 3} I_0 J_1 - {10 \over 3} I_1 J_0 \right) \; .
\ee

It remains to evaluate the sums (\ref{zs}). This is done in
Appendix A. There the term independent of temperature, which is
contained in each frequency sum (see \ref{frsum})), is neglected from
the outset. After renormalization, and if we may apply an argument
of BP (\S 2) in the present case, there remain only terms which
are down by two powers of $g$. Hence, all of the following integrals
are UV-controlled by Bose functions:
\bea \label{z12}
Z_0 &=& -2 Z_1 \;\; , \quad Z_1 \; = \; {1 \over 16 \pi^4}
      \int_{q^\ast}^\infty \! dp \, n(p) \int_{q^\ast}^\infty \! dk
      \, n(k) \; \ln \left( {p+k \over p-k} \right) \;\; , \nonu \\
Z_2 &=& -I_0 J_0 -{1 \over 2} Z_1 + {1 \over 32 \pi^4}
   \int_{q^\ast}^\infty \! dp \, n(p) \int_{q^\ast}^\infty \! dk \,
   n(k) \; {p \over k} \, \ln \left( {k^2 \over p^2-k^2} \right)
   \;\; , \nonu \\
Z_0^\prime &=& Z_1^\prime \; = \; 0  \;\; ,
\eea
where the logarithm is understood to take the absolute value of its
argument. For the delicate object $Z_2^\prime$, which had cancelled
in (\ref{hsum}), we state the singular parts here:
\be \label{z2s}
 Z_2^{\prime \; \mbox{\frm sing}} = {1 \over 16\pi^4}
    \int_{q^\ast}^\infty \! dp \, n(p) \int_{q^\ast}^\infty \! dk
    \left( p n^\prime(k) \int_{-1}^1 \!\! du {1 \over 1-u}
    - {2 p \over k} n(k) \int_{-1}^1 \!\! du \left[ {1 \over 1-u}
    \right] ^2 \; \right) \; ,
\ee
where $u=\cos (\vartheta)$ and $\vartheta$ the angle between $\vc k$
and $\vc p$. The $u$-integrals diverge when the three-momenta become
parallel. If we had worked with a corresponding cutoff $\lambda$,
most probably, $\ln (\lambda )$ and $1/\lambda$ would have appeared
in place of the $u$-integrals, respectively. We identify the above
with the 'collinear singularity' studied recently \cite{bell} in
order to establish the Kinoshita-Lee-Nauenberg theorem \cite{kino} in
thermal field theory or even in hot QCD \cite{BPS}. To justify our
identification note that the above singularity (a) occurs in a
separate factor, (b) stems from loop self-energy insertions (diagrams
7, 8, 11, 12), (c) has nothing to do with IR or UV, (d) cancels among
different contributions and, once more, (e) occurs when $\vc p$ and
$\vc k$ (or $-\vc k$) become parallel.

It remains to list the results of Appendix A for the single sums:
\bea  \label{ss}
J_0 &=& {1 \over \pi^2} \int_{q^\ast}^\infty \! dk \, n(k) \,
        {k \over 4k^2-Q_0^2} \quad , \quad J_1 \;
     = \; - {3 \over 4} J_0 \;\; , \nonu \\
I_1 &=& {T^2 \over 24} + {Q_0^2 \over 4} J_0 \quad , \quad J_0^\prime
        \; = \; J_0^{Q_0=0} + { n(q^\ast ) \over 4 \pi^2} \;\; .
\eea
These relations still contain the soft $Q_0$, since they allow for
shifting $q^\ast$ down to zero.

The relative order of the 2-loop contributions is $O(g^2)$ instead
of the $O(g)$ in search. To see this, we return to (\ref{hsum}) and
take $q^\ast$ of order $T$ in magnitude (thus allowing only for
really hard inner momenta). By substituting $p=Tp^\prime,
k=Tk^\prime$ the sums $Z_{1,2,3}$ become $T^2$ times a dimensionless
number, while $J_{0,1}$ remain to be numbers of order 1 in magnitude.
Thus, the contribution (\ref{hsum}) to $\P _\ell$ is of order
$g^4 N^2 T^2$ or, equivalently, $m^2 N g^2$ as stated above. If we
shift $q^\ast$ towards the threshold $T \wu g $, the integrands
increase and $n(k) \rightarrow T/k \,$. This amounts to at most
$O(g^{3/2} \ln (g))$ in place of $O(g^2)$. But, instead of
speculating this way, one rather should search for those other
contributions which remove the toy parameter $q^\ast$ at all. We
shall leave this point as an open question. In the present paper we
are not forced to answer it, since for the only true $O(g)$
contributions (1-loop soft) we shall see the independence of $q^\ast$
explicitly.

One might ask for any deeper reason for the null result of this
section. In general the real part of $\P$ is an even function of
$\omega$. In Appendix A, especially, odd $\omega$-powers are removed
by the operation $S_\omega \,$. Naturally, $\P$ should be considered
as a function of $\omega^2$. On the other hand, any sum $Z$ must have
the form $T^2 f(\omega^2/T^2 , \, T/m\,)$ with a dimensionless
function $f$. For hard-hard terms the second argument is absent (but
it is present at 1-loop soft). Thus, the only way to get $O(g)$ is
that $f(\omega^2/T^2)$ develops a root-singular dependence of its
argument. This is not very probable. And it did not happen indeed.
%
%
\let\dq=\thq \renewcommand{\theequation}{4.\dq}
\setcounter{equation}{0}

\parag {4. \ One-loop diagrams with hard inner momentum }
As is well known, the 1-loop contributions constitute the leading (or
'zeroth') order $\omega^{(0)} =m$ of the plasma frequency. But upon
subtracting those 'strictly hard' contributions, which are really
used to build up $m$ (and which are given an upper index zero in the
sequel), contributions of the relative order $O(g)$ might remain. In
this section we thus concentrate on that possible origin of $O(g)$
which is second in the list of next-to-leading contributions given
by BP (\S 4.3).

The contributions to $m$ are hard as well as soft. Those of $O(g)$,
if soft, are to be calculated with dressed propagators and vertices.
Thus, in sorting contributions, one is faced with an eight-fold
variety of indices: hard/soft, bare/dressed, with/without upper
index $0$. We will help ourselves by a proper definition of the term
"1-loop hard" and thereby separating it from the term "1-loop soft".

There are three 1-loop diagrams: the loop (l), the tadpole (t) and
the ghost-loop (g). Let "ltg" stand for the sum of these diagrams.
A lower index "dressed" requires to use both effective propagators
and vertices. Diagrammatically, our classification is:
\bea \label{1lh} \hspace{-1cm}
 \mbox{1-loop hard} & \equiv & \mbox{ltg (all }K)_{\,\rm bare} \;
   - \; \left[ \; \mbox{ltg (all }K)_{\,\rm bare} \; \right] ^{\, 0}
   \;\; , \\   \label{1ls} \hspace{-1cm}
 \mbox{1-loop soft} & \equiv & \mbox{ltg (soft }K)_{\,\rm dressed}
   \; - \;  \mbox{ltg (soft }K)_{\,\rm bare}   \;\; .
\eea
We now read (\ref{1lh}), (\ref{1ls}) as the specification of the
contributions $\Delta \P _\ell^{\rm 1\mbox{-}loop \; hard}$ and \\
$\Delta \P _\ell^{\rm 1\mbox{-}loop \; soft}$ to $\P_\ell$. The
prefix $\Delta$ always indicates that something is subtracted which
was already counted. Consider the second line first. If we relax the
restriction to soft $K$, (\ref{1ls}) remains still valid because,
with increasing $K$, the dressed vertices and propagators turn into
bare ones automatically. Independence of any threshold $q^\ast$ is
thus inherent in the definition. This independence (or
'UV-convergence') can be used to test intermediate results. If we
omit the $K$-specifications and add (\ref{1lh}) to (\ref{1ls}), only
one subtraction is left, namely that of the zeroth order. In passing,
the ghost-loop drops out in (\ref{1ls}) because it is not dressed.

We return to the line (\ref{1lh}) and consider the first term with
respect to $\P_\ell$ in covariant gauges. With reference to the three
diagrams l,t,g and with the notations $\Delta_0 = 1/P^2$,
$\Delta_0^- = 1/(P-Q)^2$, the result reads:
\bea \label{loop}
 \P_{\ell \; , \; \rm bare}^{\;\rm l} &=& {1 \over 2} g^2 N \sum
    \left( 2 \Delta_0 + 2 (\alpha -1) \Delta_0 + {2 \over 3}
   (\alpha -1) p^2 \Delta_0^2 - {10 \over 3} p^2 \Delta_0^- \Delta_0
   + 4 m^2 \Delta_0^- \Delta_0     \right. \nonu \\
 & & \left. \hspace{-2cm} {}
   + 4 (\alpha -1) m^2 \Delta_0^- \Delta_0 + {20 \over 3} (\alpha
   - 1) m^2 p^2 \Delta_0^- \Delta_0^2 - {1 \over 3} (\alpha -1)^2
   m^4 p^2  ( \Delta_0^- \Delta_0 )^2  \right) \;\; , \\
   \label{tadpole}
 \P_{\ell \; , \; \rm bare}^{\;\rm t} &=& {1 \over 2} g^2 N \sum
    \left( -6 \Delta_0 -2 (\alpha -1) \Delta_0 - {2 \over 3}
    (\alpha -1) p^2 \Delta_0^2 \right) \;\; ,  \\
   \label{ghost}
 \P_{\ell \; , \; \rm bare}^{\;\rm g} &=& {1 \over 2} g^2 N \sum
      {2 \over 3} p^2 \Delta_0^- \Delta_0 \;\; .
\eea
Writing down (\ref{loop}) we started with 'strictly hard' terms (the
first three), which are not at all influenced by an upper index $0$.
The tadpole contribution (\ref{tadpole}) is made up of only such
terms. It thus drops out in (\ref{1lh}). The fourth term of
(\ref{loop}) survives under the $0$-operation, but the remaining
terms (containing $m^2$) do not. The $0$-operation amounts to
$Q \rightarrow 0$ {\sl after} the frequency sum in $\sum$ has been
performed. To indicate this we write $[\,\Delta_0^-\, ]^{\, 0}
= \Delta_0^\bullet$ (instead of $\Delta_0$). With (\ref{loop}),
(\ref{tadpole}), (\ref{ghost}) the difference (\ref{1lh}) reads
\bea \label{pi1lh}
 \Delta \P_\ell^{\; \rm 1\mbox{-}loop \; hard} &=&
    {1 \over 2} g^2 N \sum \bigg( - { 8 \over 3} p^2
    (\Delta_0^- \Delta _0 - \Delta_0^\bullet \Delta_0 )
    + 4 m^2 \Delta_0^- \Delta_0  \nonu \\   & &
 + \; \mbox{the last three terms of (\ref{loop}) } \;\; \bigg) \;\; .
\eea
Let us switch to Feynman gauge for a moment. With (\ref{zs}) and with
(\ref{ss}) at $Q_0^2=m^2$ we obtain
\be   \label{pif}
  \Delta \P_{\ell \, , \;\alpha=1}^{\; \rm 1\mbox{-}loop \; hard}
  = {1 \over 2} g^2 N \left( -{3 \over 3} \left[ I_0 - I_0^\bullet
  \right] + 4 m^2 J_0 \; \right) \; = \; m^2 {5 \over 3} g^2 N J_0
  \;\; .
\ee
Hence, the magnitude of interest is hidden in
\bea   \label{jhidd}
J_0 &=& {1 \over \pi^2} \int_0^\infty \! dk \, n(k) { k \over
 4k^2-m^2 } \; = \; {1 \over 4 \pi^2} \, {\cal P}\! \int_0^\infty \!
 dx \; {x \over x^2-1} \, {1 \over e^{\varepsilon x} - 1 } \;\; , \\
& & \mbox{where} \; \varepsilon= m/2T = g \wu N / 6 \;\; , \;\;\;
    {\cal P} \; \mbox{for principal value} \;\; , \nonu
\eea
and where (A.3) has been used at $q^\ast =0$ and $Q_0^2=m^2$. At
first glance $J_0$ seems to be of order $1 / g$. However, the
integral with $1/\varepsilon x $ in place of the Bose function
vanishes. In fact, the first two terms of the asymptotic expansion of
$J_0$ are
\be \label{jasy}
   J_0 \; \sim \; {1 \over 8 \pi^2} \ln \left( {\varepsilon \over 2}
      \right) + {1 \over 4 \pi^2} \int_0^\infty \! dx \; {1 \over x}
    \left[ {1 \over e^x -1}  -  {2 \over 2x + x^2 } \right] \qquad
    (\;\varepsilon \rightarrow +0 \; )  \;\; . \qquad
\ee
Thus, $J_0$ is large only as $\ln (g)$, and $\Delta
\P _\ell^{\rm 1\mbox{-}loop \; hard}$ only reaches the relative order
$O(g^2 \ln (g))$ instead of a pure $O(g)$ in search. The terms of
(\ref{pi1lh}), which depend on $\alpha$, are also of the order of
$J_0$ (at most).

The result (\ref{jasy}) comes most opportunely, for otherwise there
would have been a dilemma. BP show that the 1-loop soft terms form a
separate gauge independent set, and argue that consequently the set
of other $O(g)$-terms must do so aswell. Thus, after we got no $O(g)$
from 2-loop hh, the 1-loop-hard terms, if $O(g)$, would have to be
gauge independent. But, according to (\ref{pi1lh}), they do depend
on $\alpha$. By the smallness of $J_0$ this is of no concern.

Once we have learned that, within the order $O(g)$, the line
(\ref{1lh}) gives zero, we may proceed simplifying the subtraction
term in the 1-loop-soft line, which is the last term of (\ref{1ls})
(without the minus). There, the 'total hard' parts of (\ref{loop})
and (\ref{tadpole}) suffice. We will write down the subtraction terms
for loop and tadpole separately. The gauge dependent pieces of these
two terms cancel (reflecting the gauge independence of the zeroth
order). But when kept separately, these terms (e.g. the last two in
(\ref{tadpole})) could be (and are indeed) necessary to restore
$q^\ast$-independence. However, as all $\alpha$'s will drop out in
the sequel before UV details need be studied, we need not keep them.
The subtraction terms are now prepared as
\bea \label{lsub}
 \P_{\ell \; , \; \rm bare}^{\;\rm l} &=& g^2 N \sum \left(
    \Delta_0 - {5 \over 3} \Delta_0^- \Delta_0 \, p^2 \,\right) \\
 \label{tsub}
 \P_{\ell \; , \; \rm bare}^{\;\rm t} &=& g^2 N \sum \left(
     - 3 \Delta_0 \right)
\eea
for use in the following section. (\ref{lsub}) and (\ref{tsub}) are
real. Therefore, if one studies imaginary parts \cite{6.6}, no
subtractions are required. Gauge dependences however may remain in
the dressed tadpole and the dressed loop. We shall check their
cancellation.
%
%
\let\dq=\thq \renewcommand{\theequation}{5.\dq}
\setcounter{equation}{0}

\parag {5. \ One-loop diagrams with soft inner momentum }

\noindent \parbox{16cm}{5.1 \ THE TADPOLE DIAGRAM \vspace{.7cm} }
\hfill \vphantom{a} \nopagebreak \indent
As discussed in the preceding section there are precisely two
diagrams, tadpole and loop, which might (and do) contribute at order
$O(g)$ through (\ref{1ls}). In this subsection we concentrate on
the first. Using the dressed 4-vertex (\ref{gam4}) and the dressed
gluon (\ref{gmn}) the first term in (\ref{1ls}) becomes
\be \label{5pimn}
 \P\omn_{\rm tadpole} = {1 \over 2} g^2 N \sum_P^{\rm soft}
  G(P)_{\lambda \, \rho} \left( \left[ g^{\mu \rho} g^{\lambda \nu}
  + g^{\mu \lambda} g^{\rho \nu} - 2 g^{\mu \nu} g^{\rho \lambda}
  \right] + \delta \Gamma^{\mu \nu \rho \lambda} (Q,-Q,-P,P) \right)
\ee
This is equation (4.23) of BP (apart from the sign, which is
notational). Using the relation $A+B+D=g$, the propagator
may be written as
\be  \label{5g}
 G = g \; \Delta_\ell + D \; \Delta_{\alpha \, \ell}
    + A \; \Delta_{t \, \ell}  \quad  \mbox{with} \quad
     \Delta_{\alpha \, \ell} \equiv \Delta_\alpha - \Delta_\ell
     \;\; \mbox{etc.} \; .
\ee
Turning to the longitudinal part $\P_\ell$ of (\ref{5pimn}) by taking
the trace with $B$, i.e. by sandwiching (\ref{5g}) with vectors $V$
and dividing by $V^2$, we may decompose into three parts as follows
\be \label{5pis}
 \P_\ell^{\rm tadpole} = \P_{\rm bv} +\P_{\alpha\,\ell} + \P_{t\,\ell}
\ee
The index ${\rm bv}$ refers to the bare vertex in (\ref{5pimn}). The
other two terms combine the HTL-part of the vertex with the $\alpha
\ell$- and $t\ell$-part of the propagator. Its first term,
$g \Delta_\ell$, may be omitted, because it does not contribute to the
order $O(g)$ under consideration (it leads to an integral $J_1$ and
only reaches the order of those terms already neglected in section 3).

The quantity $\P_{\rm bv}$ is easily evaluated. Note that the matrices
$B$, $A$, $D$ occur at two different arguments. This amounts to
 \[
  \Tr B(Q) A(P) \rightarrow 2/3 \quad , \quad \Tr B(Q) D(P)
  \rightarrow - p^2 / 3 P^2 \qquad ( \vc q \rightarrow 0 )
 \]
when the limit $\vc q \rightarrow 0$ is taken. In this limit we obtain
\be \label{5dbv}
\Delta \P_{\rm bv} = g^2 N \sum_P^{\rm soft}\left( -3 \Delta_{\ell\, 0}
  - {4 \over 3} \Delta_{t \, \ell} - \Delta_{\alpha \ell}
  - {1 \over 3} \Delta_{\alpha \ell} \, \Delta_0 \, p^2 \right) \;\; .
\ee
In writing down (\ref{5dbv}) we have included the subtraction
term (\ref{tsub}), which is the second in $\Delta_{\ell \, 0}
= \Delta_\ell - \Delta_0$. By this subtracion the first term
of (\ref{5dbv}) becomes UV convergent. Nevertheless,
the index 'soft' on the sum in (\ref{5dbv}) remains
necessary in order to control the two $\alpha$-dependent terms.

With (\ref{5dbv}) we have reached a point where one can for the first
time see how a true $O(g)$-term comes about in a natural manner.
So, let us stay with the $\Delta_{\ell \, 0}$-term of (\ref{5dbv})
and evaluate. The propagators we work with (see e.g. (\ref{proptab})
below) are even functions of $P_0$ and of $p$. Hence, their spectral
densities $\rho (x,p)$ are odd functions of $x$:
\be  \label{5spect}
 \Delta (P) = \int \! dx \, { \rho (x,p) \over P_0 - x } =
 \int \! dx \, x {\rho (x,p) \over P_0^2 -x^2 } \;\; .
\ee
To calculate $\sum \Delta$ we need
\be  \label{5frsum}
 \sum_{P_0} { 1 \over P_0^2 -x^2 } = - {1 \over 2 \pi i}
 \int_{\!\bigcirc} dP_0 \, {n(P_0 )\over P_0^2 -x^2}
  = - {1 + 2 n(x) \over 2 x } \rightarrow - {T \over x^2} \;\; ,
\ee
where the integral surrounds the whole complex $P_0$-plane
counterclockwise. If $p$ is soft also $x$ is soft since otherwise
the density $\rho$ vanishes. Therefore, the leading term of
(\ref{5frsum}) can be extracted as shown to the right. For any
propagator $\Delta$ with such properties we thus have
\be   \label{5singsum}
  \sum^{\rm soft} \Delta \, f(p) = - T \int_P^3 f(p) \, \psi (p)
  \quad \mbox{with} \quad
  \psi (p) \equiv \int \! dx \, {1 \over x} \rho (x,p) \;\; ,
\ee
where $f$ is an arbitrary weight function. In the case at hand we have
$f=1$, and the minus-first moment $\psi _\ell (p)$ can be taken from
the table (\ref{proptab}) or from Appendix B:
\be \label{5L}
 - \sum^{\rm soft} \Delta_\ell = T \int_P^3 \psi_\ell (p)
   = {T \over 2 \pi^2} \int_0^{q^\ast} \! dp \, {p^2 \over 3m^2+p^2 }
     \; \equiv \; {\cal L} \;\; .
\ee
${\cal L}$ is UV-divergent, i.e. it depends on $q^\ast$. The sum
$\sum \Delta_0$, if evaluated just so, has the same sort of
divergence. The difference is finite:
\bea \label{5K}
 - \sum \Delta_0 &=& {T \over 2 \pi^2} \int_0^{q^\ast} \! dp \qquad ,
   \qquad \sum \Delta_{\ell \, 0} \; = \; 3 {\cal K}
   \qquad \;\; \mbox{with} \nonu \\
  {\cal K} &\equiv &  - \sum {m^2 \over p^2 } \Delta_\ell \; = \;
  {T \over 2 \pi^2} \int_0^\infty \! dp \, {m^2 \over 3m^2+p^2 }
       \; = \; {m T \over 3} \; { \wu 3 \over 4 \pi } \;\; .
\eea
Thus, the considered $\P$-contribution is of the order
$g^2 N m T = m^2 \, 3 g \wu N $ in magnitude. (\ref{5K}) shows that
the 'odd power' $g$ arises via a simple substitution thanks to the
presence of the scale $m$. Our expressions at 2-loop hard-hard and
1-loop hard had no such scale. We learn that there are terms of order
$O(g)$, indeed. Furthermore, sums over single propagators can be
evaluated (for a collection see Appendix D). But, as a rule, sums
over pairs of propagators (at different arguments) need numerical
evaluation.

The remaining two terms in (\ref{5pis}) are more complicated. They
both represent examples for the excellent utility of Ward identities.
We start, using vertical vector notation, from
\be  \label{5verti}
 \P_{ \left\{ \vphantom{a}^{\alpha \; \ell}_{t \; \ell} \right\} }
    = {1 \over 2} g^2 N {V_\mu V_\nu \over V^2}
      \sum_P^{\rm soft} \left\{ \matrix{\Delta_{\alpha \ell} (P)
   D_{\lambda \rho} (P) \cr \Delta_{t \ell} (P) A_{\lambda \rho}
   (P) \cr} \right\} \delta \Gamma^{\mu \nu \lambda \rho} (Q,-Q,-P,P)
  \;\; .
\ee
In the upper line $D (P) = P \circ P / P^2$ brings in the momenta
the Ward identities (\ref{ward4}) and (\ref{ward3}) are formulated
with. Using both one obtains:
\be \label{5ward}
  P_\lambda P_\rho \delta\Gamma^{\mu \nu \lambda \rho} (Q,-Q,-P,P)
  = 2 \P\omn (P-Q) - 2 \P\omn (Q)  \;\; ,
\ee
where use has been made of the fact that the above $\P\omn$ is an
even function of its argument and so is $\Delta_{\alpha \ell}$. In
the difference (\ref{5ward}) we take care to do the same mainpulations
on both terms. Using (\ref{pino}), sandwiching with $V$-vectors,
replacing $(VK)^2/V^2$ by $- \vc k ^2 /3$ at $\vc q \rightarrow 0$
and forming a common denominator (which is $K^2 (K-Q)^2 (K-P)^2$; the
numerator is $2KP-P^2$ and the term $P^2$ may be neglected), a hard
$K$-sum is obtained. It is given in Appendix C (form $\vartheta +
\varphi$ there). This leads to the final form of the term studied:
\be   \label{5alend}
 \P_{\alpha \ell} ={1 \over 3} g^2 N \sum_P^{\rm soft}
 \Delta_{\alpha \ell} \P_\ell (P-Q) \Delta_0 \Delta_0^- p^2
\ee

In the lower component of (\ref{5verti}), appearently, there is no
momentum in front of $\delta \Gamma$ as is needed in the Ward
identity. We can produce such momenta, however, by the following
exotic line which works at $\vc q = 0$:
\be   \label{5exo}   \hspace{-.4cm}
   - 3 {V_\mu V_\nu \over V^2 } K^\mu K^\nu \rightarrow \vc k ^2
   =  {(KQ)^2 - K^2 Q^2 \over Q^2 } = \left( {Q_\mu Q_\nu \over Q^2 }
   - g\mn \right)  K^\mu K^\nu \approx {Q_\mu Q_\nu \over Q^2}
     K^\mu K^\nu \; .
\ee
The $g\mn$-term may be neglected, if (\ref{5exo}) is used with two
of the $K$'s in $\delta \Gamma^{1234}$ (typically, such terms can be
neglected when deriving the Ward identity (\ref{ward4})).
Note that $\delta \Gamma$ is invariant under the interchange
of Q with P. Thus, with (\ref{ward4}) and (\ref{ward3}), we have
\be \label{5qqg}
   Q_\mu Q_\nu \delta \Gamma^{\mu \nu \lambda \rho} (Q,-Q,-P,P)
 = Q_\mu Q_\nu \delta \Gamma^{ \lambda \rho \mu \nu}(P,-P,-Q,Q)
 = 2 \P^{\lambda \rho} (P-Q) - 2 \P^{\lambda \rho} (P) \;\; .
\ee
Using again (\ref{pino}) together with
\be  \label{5kak}
   - K^\lambda A_{\lambda \rho} K^\rho = k^2 -
     {(\vc k \vc p )^2 \over p^2 }  \;\; ,
\ee
which is an invariant under $K \rightarrow P-K$, we end up with the
sum $\beta$ of Appendix C. The result is
\be   \label{5ptl1}
 \P_{t \ell} = g^2 N \sum_P^{\rm soft} \Delta_{t \ell} (P) {1 \over
  3 m^2} \, \left[ \; \P_\ell (P-Q) - \P_\ell (P) \; \right] \;\; .
\ee
As the above derivation used the somewhat dangerous line
(\ref{5exo}) we like to mention, that originally $\P_{t \ell}$
and $\P_{\alpha \ell}$ were evaluated 'by hand', i.e. by
first doing the frequency sums and preparing the leading
terms afterwards. The results were indeed the same.

At the second term in (\ref{5ptl1}) we encounter the possibility to
cancel a self-energy in the numerator against the same contained in
the propagator. We shall do so whenever possible following a hint in
Ref. \cite{6.6}. As this step occurs repeatedly in the next section,
let us work with a short hand notation,
\be  \label{5ab}
   m^2 - P^2 \equiv a \quad , \quad m^2 - \P_\ell (P) \equiv b \quad ,
   \quad \Delta_\ell = {1 \over b-a } \quad , \quad
   \Delta_t = {-2 \over b+2a} \;\; ,
\ee
where the identity $\P_\ell + 2\P_t = 3 m^2$ has been exploited. Then:
\be  \label{tlpil}
\Delta_{t \ell } \P_\ell \; = \;  \left( {-2 \over b+2a}
    - {1 \over b-a} \right) \left( m^2-b \right)  \; = \;
      m^2 \Delta_{t \ell} + 2 \Gamma_t + \Gamma_\ell \;\; ,
\ee
where by $\Gamma \equiv 1 + a \Delta$ we keep terms together such
that the behaviour as $1 / P^2$ at large $P$ may be associated with.
Due to their neat properties we call $\Gamma_\ell$ and $\Gamma_t$
'propagators'. Let us list them together with other
propagators in use:
\def\ms{ \vspace{.14cm} \\ }
\bea  \label{proptab}
\hspace{-.4cm} \begin{array}{*{4}{c@{\;\;}}}
\mbox{propagator} & \mbox{spectral density} \; \rho
                        & \mbox{relation} &      \psi (p)          \ms
\Delta_0    & \dis {1\over 2p} \left[ \,\delta (x-p) - \delta(x+p)
              \,\right] & \dis \Delta_0 = {1 \over P^2}
                        & \dis {1 \over p^2}                       \ms
\Delta_t    & \rho_t    & (\ref{gmn}) & \dis {1 \over p^2 }        \ms
\Delta_\ell & \rho_\ell & (\ref{gmn}) & \dis {1 \over 3 m^2 + p^2 }\ms
\Gamma_t    & \rho_t \; (m^2+p^2-x^2 ) & \Gamma_t = 1 +
                   (m^2 -P^2) \Delta_t  & \dis { m^2 \over p^2 }   \ms
\Gamma_\ell & \rho_\ell \; (m^2+p^2-x^2) & \Gamma_\ell = 1 +
           (m^2-P^2)\Delta_\ell   & \dis {-2m^2 \over 3m^2+p^2 }   \ms
\Omega_t & \rho_t \; (x^2-p^2)
         & \Omega_t = m^2 \Delta_t - \Gamma_t & 0                  \ms
\Omega_\ell & \rho_\ell \; (x^2-p^2) & \Omega_\ell = m^2 \Delta_\ell
              - \Gamma_\ell      & \dis {3m^2 \over 3m^2+p^2}      \ms
\Delta_\theta & \dis {3 m^2 x \over 2p} \, {p^2 - x^2 \over p^2} \,
   \theta (p^2-x^2) & \Delta_\theta = m^2-\P _\ell (P) & \dis 2m^2
\end{array}  \eea
\smallskip \noindent
The argument of each propagator is $P$. The object $\Delta_\theta$
is nothing but the quantity $b$ in (\ref{5ab}). It is a 'propagator'
having no pole contribution. According to (\ref{5spect}), in each
case the minus-first moment $\psi$ equals minus the propagator at
$P_0=0$.

To apply the above reformulations, the third term of the soft
tadpole contribution (\ref{5pis}) is given by
\be    \label{5tlend}
\P_{t \ell} = - g^2 N {1 \over 3 m^2} \sum_P^{\rm soft}
  \, \left[ \; 2 \Gamma_t + \Gamma_\ell +
  \Delta_{t \ell}^- \Delta_\theta \; \right] \;\; .
\ee
One may require that the expression (\ref{5tlend}) is UV-stable
automatically and needs no cutoff. This is indeed the case.
Consider for example $2 \Gamma_t + \Gamma_\ell$. As the table
(\ref{proptab}) shows, the moment $\psi$ of this combination behaves
as $p^{-4}$ at large $p$ as required.

To summarize this subsection we evaluate the sums over single
propagators in (\ref{5dbv}) and (\ref{5tlend}), see Appendix D, and
obtain
\be  \label{tadend}
  \Delta \P_\ell^{\rm tadpole} = g^2 N \left( -3 {\cal K}
    - {1 \over 3m^2} \sum_P \Delta_{t \ell}^- \Delta_\theta \right)
    + \quad \matrix{ _{{\rm terms\; containing}\; \Delta_{\alpha \ell}
           \, ,\; {\rm which\; will}} \cr
          ^{\rm cancel\; against\; those\; in\; the\; loop } \cr } \;\; .
\ee

\bigskip
\noindent 
\parbox{16cm}{5.2 \ THE LOOP DIAGRAM  \vspace{.7cm} }
\hfill \vphantom{a} \nopagebreak \indent
The loop has the symmetry factor 1/2. It is made up of two dressed
3-vertices (\ref{gam3}) and two dressed gluons. After the colour sums
are done the contribution reads
\be \label{loopstart}
    \P\omn_{\rm loop} = {1 \over 2} g^2 N \sum_P^{\rm soft}
    G(P-Q)_{\rho \sigma} G(P)_{\lambda \tau} \Gamma (Q,P-Q,-P)^{\mu
    \sigma \tau} \Gamma(Q,P-Q,-P)^{\nu \rho \lambda} \;\; .
\ee
Turning to the longitudinal part, each vertex is contracted with
one $V$-vector. Since each propagator $G$ is made up of the
three terms (\ref{5g}), which we now number from 1 to 3, nine
contributions can be distingushed:
\be  \label{rij}
  \P_\ell^{\; \rm loop} = {1 \over V^2} V_\mu \P\omn_{\rm loop}
  V_\nu \equiv \sum_{i,j=1}^3 \P_{ij} \quad , \quad \P_{ij}
  = {1 \over 2} g^2 N \sum_P^{\rm soft} {\cal R}_{ij} \;\; .
\ee
An element $\P_{ij}$ or ${\cal R}_{ij}$ depends on $\alpha$, if
at least one index takes the value 2. Thus, if (\ref{rij}) is
arranged as a $3\times 3$-matrix, the gauge dependent
elements form a 'red cross'. To comment the next steps consider
for example ${\cal R}_{23}$ :
\be  \label{r23}
 {\cal R}_{23} = \Delta_{\alpha \ell}^- \Delta_{t \ell}
    {1 \over (P-Q)^2 } {1 \over V^2}\, W^\lambda A_{\lambda \tau}
    W^\tau \quad , \quad \mbox{with}  \quad W^\lambda \equiv
    V_\mu (P-Q)_\rho \Gamma^{\mu \rho \lambda} \;\; .
\ee
Using the details of $\Gamma^{123}$, (\ref{gam3}), one obtains
\bea   \label{waw}  \hspace{-.8cm}
  WAW &=& (P^2-Q^2)^2 \left( V^2+ {(VP)^2 \over p^2 } \right)
            \nonu \\
& & \hspace{-2cm} {} - 16 g^2 N (P^2-Q^2)\sum {1\over N_K } (KP-KQ)
    \left( (VK)^2 - (VK)(VP) {\vc k \vc p \over p^2 } \right) \nonu \\
& & \hspace{-2cm} {} - (8g^2N)^2 \sum \sum
     {(KP-KQ)(RP-RQ) \over N_K N_R } (VK) (VR) \left( \vc k \vc r -
     {(\vc k \vc p ) (\vc r \vc p ) \over p^2} \right) \; ,
\eea
where $N_K = K^2 (K-Q)^2 (K-P)^2 $. The expression simplifies
slightly through $\vc q \rightarrow 0$ since e.g.
$(VK) (VP) / V^2 \rightarrow - (\vc q \vc k ) (\vc q \vc p ) / q^2
\rightarrow - \vc k \! \vc p  / 3 $, where the last step exploited
the angular integrations contained. Note that $Q=(Q_0,\vc 0 )$ in all
what follows. The double sum, which runs over $K$ and $R$, both hard,
can be decoupled, see Appendix C. This decoupling is possible in all
nine cases and leads to squares of various single sums (over
hard $K$ and soft outer variable), which can be evaluated towards
the leading term. They are listed in Appendix C and denoted by small
greek letters. It should be noted that, most probably, all we do
at this stage was already worked out by BP in \S 4.2. But let us
be obstinate in order to have an independent test.

We leave the above special example and notice the result for all nine
terms at the same intermediate level. The element ${\cal R}_{22}$ in
the center of the 'red cross' is the only one containing squares of
$\alpha$:
\be  \label{r22}
 {\cal R}_{22} = - {1 \over 3} \Delta_{\alpha \ell}^-
        \Delta_{\alpha \ell} \Delta_0 \Delta_0^- p^2
         \left[ Q^2 - 8g^2N \tau \right] ^2
\ee
The object $\tau$ is the hard $K$-sum mentioned above. Its
evaluation is straightforward:
\be \label{tau}
   \tau \equiv \sum {(KP) (KQ-KP) \over N_K }\, { \vc k \vc p \over
   p^2} \quad , \quad
   8 g^2 N \tau = m^2 + \; \mbox{non-leading terms} \;\; . \quad
\ee
Using (\ref{tau}) the square bracket in (\ref{r22}) becomes $Q^2-m^2$.
Thus the whole term vanishes on mass shell. No $\alpha^2$ survives.
The other gauge dependendent pieces stand at the four ends of
the 'red cross':
\bea   \label{rends}
 {\cal R}_{12} &=& \Delta_\ell^- \Delta_{\alpha \ell} \Delta_0
             \Bigg( [(P-Q)^2 - Q^2 ] ^2 + {p^2 \over 3}
            \left( P^2 -2PQ-Q^2 \right)  \nonu \\
    & & {} - {16 g^2N\over 3} \left[ (2PQ-P^2)(\vartheta + \varphi)
      + \tau \right] - {1 \over 3} (8g^2N)^2 \left[ {\lambda ^2 \over
    p^2 } -{\vartheta ^2 \over 2} - \varphi ^2 \right] \Bigg) \nonu \\
 {\cal R}_{23} &=& {1 \over 6} \Delta_{\alpha \ell}^- \Delta_{t \ell}
     \Delta_0^- \left( 2 P^2 - 2Q^2 + 8g^2N \mu \right) ^2   \nonu \\
  {\cal R}_{21} &=& {\cal R}_{12} \quad , \quad
  {\cal R}_{32} \; = \; {\cal R}_{23} \;\; ,
\eea
where the hard $K$-sums read $\vartheta$, $\varphi$, $\tau$, $\lambda$,
$\mu$ and are defined in Appendix C. Their leading terms are:
\be    \label{grrel1}
 8g^2N\varphi = m^2 - \P_\ell^- - p^2 \Delta_0^- \P_\ell^-
 \;\; , \;\; 8g^2N\mu = \P_\ell -m^2 \;\; , \;\; \vartheta = \mu^-
 \;\; , \;\; \lambda = (P_0-Q_0) (\varphi + \vartheta) \;\; .
\ee
In passing, the symmetry ${\cal R}_{ij} = {\cal R}_{ji}$ arises
only when the leading parts of the $K$-sums are taken. In
general, if $i \neq j$, ${\cal R}_{ji}$ differs from
${\cal R}_{ij}$ in the denominators, which are $K^2 (K+Q)^2 (K+Q-P)^2$
in place of $N_K$.

It is irresistible to look back at the gauge dependent pieces of the
tadpole contribution: (\ref{5alend}) and the last two terms in
(\ref{5dbv}). Both contain $\Delta_{\alpha \ell}$, and the expressions
(\ref{rends}) do so as well (note that $P\rightarrow Q-P$ is allowed
in any ${\cal R}$-element). To rewrite ${\cal R}_{23}$ we set
$Q^2=m^2$, insert (\ref{grrel1}) and use the notation (\ref{5ab}):
\be
 {\cal R}_{23} = - {1 \over 3} \Delta_{\alpha \ell}^- \Delta_0^-
          (b+2a) - {1 \over 6} \Delta_{\alpha \ell}^-
          \Delta_\ell \Delta_0^- (b+2a)^2  \;\; .
\ee
For ${\cal R}_{12}$ we observe that it is the sum of three squared
brackets containing one 'greek' sum each. The next steps are
$P\rightarrow Q-P$, inserting (\ref{grrel1}), using again the
$a$-$b$-notation, but still maintaining the order of the
three squared 'greeks' of (\ref{rends}):
\be
 {\cal R}_{12} = {1 \over 3} \Delta_{\alpha \ell}^- \Delta_\ell
  \Delta_0^- \left( (b-a)^2 \Delta_0^2 P_0^4 + {1\over 2} (b+2a)^2
  - (b-a)^2 \Delta_0^2 P_0^2 p^2 \right)  \;\; .
\ee
The second term will cancel the last term of ${\cal R}_{23}$.
and one $(b-a)$ suffices to kill $\Delta_\ell$. After a few
rather trivial steps (including the omission of a term which
is odd in $P$) we end up with
\be
 {\cal R}_{12}+{\cal R}_{21}+{\cal R}_{23}+{\cal R}_{32} =
  {2 \over 3} \Delta_{\alpha \ell} \Delta_0 \left(
  3P^2 + p^2 - \Delta_0^- \, p^2 \, \P_\ell^-  \right) \;\; .
\ee
Obviously, these three terms precisely cancel the gauge dependence
parts of the tadpole. BP are right. The order $O(g)$-terms form a
gauge independent set. As this result was expected (see the text
below (\ref{logg})), it merely tells us that the procedure followed
so far works smoothly.

The physics is contained in the four ${\cal R}$-elements in the
corners of the matrix:
\bea  \label{physr}
 {\cal R}_{11} &=& \Delta_\ell^- \Delta_\ell \bigg( 2P^2-2PQ+5Q^2
   - {10 \over 3} p^2 - {16 g^2N \over 3}
   \left[ (Q_0-2P_0)\, (\beta + \gamma) + 2 \sigma \right] \nonu \\
  & & {} - {1 \over 3} (8 g^2 N)^2 \left[ {\sigma^2 \over p^2}
         - \beta^2 -2\gamma^2 + {3 \rho^2  \over 2 p^2} +
         {\zeta^2 \over p^2 } \right] \bigg)   \nonu \\
 {\cal R}_{13} &=& \Delta_\ell^- \Delta_{t \ell} \bigg( {2\over 3}
         (P+Q)^2  - {8 p^2 \over 3} + {16 g^2N\over 3}
         \left[ (P_0+Q_0) \beta + \rho \right]
         + {1 \over 3 } (8g^2N)^2 \left[ {\beta ^2 \over 2}
         - {\rho^2 \over p^2 } \right] \bigg)   \nonu \\
 {\cal R}_{31} &=& {\cal R}_{13}  \nonu \\
 {\cal R}_{33} &=& - \Delta_{t \ell}^- \Delta_{t \ell}
    {1\over 6 p^2} \left( 4 p^2 -8g^2N \rho \right) ^2 \;\; .
\eea
The hard $K$-sums contained here, which are again defined in Appendix
C, can be traced back to those already given in (\ref{grrel1}):
\be  \label{grrel2}
 Q_0 \beta= \mu^- - \mu \;\; , \;\; Q_0 \gamma = \varphi - \varphi^-
   \;\; , \;\; Q_0 \sigma = \lambda^- - \lambda \;\; , \;\; \rho =
   P_0 \beta - \vartheta \;\; , \;\; \zeta = \sigma - \rho \;\; .
\ee

In the sequel the above ${\cal R}$-elements are subject to several
transformations and regroupings with the general aim of
simplification. For instance, we cancel self-energies in the numerator
as in (\ref{tlpil}) and reduce $P$-powers by changing from $\Delta$-
to $\Gamma$-propagators. The procedure ends up with the four standard
expressions given in (6.1), (6.2) below. To illustrate the steps we
concentrate on the derivation of ${\cal M}_2$, see (6.2). This term is
part of ${\cal R}_{33}$. Admittedly, this is the simplest term of
(\ref{physr}). At first we insert $\rho$, (C.9), square out and reduce
the number of terms slightly by means of $P \rightarrow Q-P$. In pure
factors $Q_0^2$ or $1/Q_0^2$ we replace $Q_0^2$ by $m^2$, thus
anticipating the analytical continuation otherwise to be done at the
end. The resulting expression
\be
{\cal R}_{33} = - {1 \over 3 m^2 p^2 } \,
     \Delta_{t \ell}^- \Delta_{t \ell} \left( 8 m^2 p^4
     + P_0 (Q_0 - P_0 )b b^- - 8 p^2 P_0 Q_0 b
     + P_0^2 b^2 \right)
\ee
is ready for cancellations of $b$ as often as possible:
\be
   \Delta_{t \ell} \, b = - 2 \Gamma_t - \Gamma_\ell  \quad , \quad
   \Delta_{t \ell} \, b^2 = - 3 b + a  \left( 4 \Gamma_t
   - \Gamma _\ell \right)  \;\; .
\ee
Clearly, there remain terms of the form $\Delta^- b$, which will be
collected at the end to give ${\cal M}_1$, see (6.2). Next we keep
only terms having the index $t$ twice and denote this selection
by ${\cal R}_{33}^{tt}\, $:
\be
 {1\over 2} {\cal R}_{33}^{tt} = - {4 \over 3} \Delta_t^-
   \Delta_t p^2 - {2 P_0 (Q_0-P_0) \over 3 m^2 p^2} \Gamma_t^-
   \Gamma_t - {8 \over 3 m^2} \Delta_t^- P_0 Q_0 \Gamma_t
   - {2 \over 3 m^2 p^2}  \Delta_t^- P_0^2 a \Gamma_t \;\; .
\ee
In the last term we use the identity
\be
 P_0^2 a = \left( 3m^2 + 2 P_0 Q_0 + P_0^2 \right) \, a^-
  - (3m^2 + p^2 ) 2 P_0 Q_0 - 3 m^2 p^2 \;\; ,
\ee
valid at $Q_0^2=m^2$, in order to reduce $P_0$-powers. Note that
$a^-\Delta_t^- = \Gamma_t^- - 1$, where the $-1$ leads to terms with
only one propagator. The product $2 P_0 Q_0 $ may be
replaced by $m^2$, if it occurs with factors symmetric under
$P \rightarrow Q-P$. Even if it occurs with $\Delta_t^- \Gamma_t$,
we may write
\[   \Delta_t^- \Gamma_t \, 4 P_0 Q_0 = 4 \Delta_t^- \, P_0 Q_0
     + \Delta_t^- \Delta_t \, 4 P_0 Q_0 a    \]
and symmetrize in the last term by
\[ 4 P_0 Q_0 a \rightarrow m^2 \left( 3 a + 3 a^-
          - 3 m^2 - 4 p^2 \right) \]
followed by $\Delta_t a = \Gamma_t - 1$. Note the two more origins
of single-propagator terms (SPT). The procedure ends up with
\be   \label{33tt}
  {1 \over 2} {\cal R}_{33}^{tt} = - 3 m^2 \Delta_t^- \Delta_t
  - 3 \left( m^2 \Delta_t^- - \Gamma_t^- \right)
      \left( m^2 \Delta_t -   \Gamma_t   \right) + {\rm SPT}
\ee
with
\bea   \label{spt}
 {\rm SPT} &=& {8 \over 3} \Gamma_t {1 \over p^2} - {2 \over 3}
     \left( \Gamma_t^- - \Gamma_t \right) {P_0\over Q_0 }
     {1 \over p^2} - 4 \Delta_t {m^2\over p^2 } + 2 \left( \Delta_t^-
     - \Delta_t \right) {P_0 Q_0 \over p^2 } \nonu \\
  & & {} + {4\over 3} \Delta_t - {2 \over 3} \left(
            \Delta_t^- - \Delta_t \right) {P_0 \over Q_0} \;\; .
\eea
For the sum over SPT we read off from Appendix D that
\be    \label{sspt}
  \sum {\rm SPT} = - {8 \over 3} {\cal V} + {2 \over 3} {\cal V}
  + 4 {\cal V} - 2 {\cal V} - 2 {\cal K} - {2 \over 3} {\cal L} \;\; ,
\ee
where the first four terms correspond to the first four in (\ref{spt}).
Each of these four sums diverges in the infrared, since
\be    \label{5V}
 {\cal V} \; \equiv \; - \sum \Delta_t {m^2 \over p^2} \; = \;
 {T m^2 \over 2 \pi^2 } \int_0^{q^\ast } dp {1 \over p^2 } \;\; ,
\ee
but they cancel each other in (\ref{sspt}). We also see how terms
denoted by ${\cal K}$ arise. They are collected in the first term of
(6.1). The UV-singular objects ${\cal L}$ either cancel or are needed
in (6.2) to compensate the $q^\ast$-dependence of the first part of
${\cal M}_1$. Clearly, the first two terms of (\ref{33tt})
give ${\cal M}_2$ in (6.2), as announced. At first glance,
the propagator $\Omega_t$ is introduced for a shorter notation only.
Note however that in the IR-region $\Omega_t$ is less dangerous
than $\Gamma_t$, as can be seen in (\ref{proptab}).

The treatment of ${\cal R}_{11}$, ${\cal R}_{13}$ and even of the
remaining parts of ${\cal R}_{33}$ along the steps just described
leads into a lenghty and tedious procedure. Here we only comment on
one more detail. After the $b$-cancellations in ${\cal R}_{11}$ are
done, a term $\Delta_0^- \, \Delta_0 \, p^2 $ is left, which also
occurs in the subtraction-term (\ref{lsub}) (but the two do not
cancel). The sum over it gives $I_1$ as defined in (\ref{zs}) and to
be evaluated there at hard inner momentum. But since the result
(\ref{ss}) allowed for $q^\ast \rightarrow 0$, we may write
\be
  \sum \Delta_0^- \Delta_0 \, p^2 = - {1 \over 2} \sum \Delta_0
     + {m^2 \over 4 } J_0
\ee
and turn to soft integration momentum. Since the $J_0$-term may be
neglected, see (\ref{jhidd}) and (\ref{jasy}), we learn that the
replacements
\be   \label{null}
  \sum \Delta_0^- \Delta_0 \, p^2 \rightarrow {1\over 2}
  {\cal L} + {3 \over 2} {\cal K} \qquad \mbox{and} \qquad
  m^2 \sum \Delta_0^- \Delta_0 \rightarrow 0
\ee
are allowed in ${\cal R}_{11}$ as well as in the subtraction term.
By including the latter this subsection ends up with
\be  \label{loopend}
  \Delta \P_\ell^{\rm loop} = g^2 N \left( \, {11 \over 6} {\cal L}
     + {11 \over 2} {\cal K} \, + \, \sum_P^{\rm soft} \, \left[ \,
 {1 \over 2} {\cal R}_{11} + {\cal R}_{13} + {1 \over 2} {\cal R}_{33}
 \, \right] \; \right) \;\; .
\ee
\newpage
%
%
\let\dq=\thq \renewcommand{\theequation}{6.\dq}
\setcounter{equation}{0}

\parag {6. \ Result }
The analysis of the preceeding sections may be summarized as follows.
The only $O(g)$-contributions to the real part of the polarisation
function $\P_\ell$ arise from the soft tadpole and the soft loop,
(\ref{tadend}) and (\ref{loopend}). Their sum is independent of the
gauge parameter $\alpha$ and may be cast into the following form:
\be    \label{magni}
\Delta \P_\ell^{\; \rm 1\mbox{-}loop \;\, soft} \; = \; g^2 N \left(
   \; 4 \, {\cal K} + \sum_{j=1}^4 {\cal M}_j \; \right)
\ee
with
\bea  \hspace{-1cm}
  {\cal M}_1 = {1 \over 6m^2} \sum \Delta_t^- \Delta_\theta
       {P^2 \over p^2} + {1 \over 3} {\cal L} & \quad , \quad &
  {\cal M}_2 = - 3 m^2 \sum \Delta_t^- \Delta_t - 3 \sum \Omega_t^-
       \Omega_t {1 \over p^2} \quad , \nonu \\
       \label{magni2}  \hspace{-1cm}
  {\cal M}_3 = - {3 \over 2m^2} \sum \Omega_\ell^-
       \Omega_t {P^2 \over p^2} \hspace{.9cm} & \quad , \quad &
  {\cal M}_4 = - {3 \over 2} \sum \Omega_\ell^- \Omega_\ell
       {1 \over p^2} \quad .
\eea
Any non-covariant gauge should lead to (6.1) as well. The first
term of (6.1), $4 {\cal K}$, is positive. But the whole
contribution is expected to be negative.

Obviously, the above four terms carry different index pairs.
But there are more properties in favour of the decomposition.
Each term ${\cal M}_j$
\def\ab#1{\\ \phantom{a} \hspace{.174cm} (#1) \,\,}
\def\eb{\\ \phantom{a} \hspace{1.2cm}}
\ab a converges at large $P_0$ when summing over frequencies
\ab b is UV-stable, i.e. it does not depend on the cutoff $q^\ast$
\ab c is IR-stable (this forces the two terms of ${\cal M}_2$
      together, see below)
\ab d contains two propagators, where one is taken at argument
      $Q-P$. Moreover, \eb each ${\cal M}$ either has the form
      $\sum \Delta^- \Delta \, f(p)$ or it can be cast into it
      (see below).

The statement (a) is a rather trivial consequence of the spectral
representation (\ref{5frsum}). The latter shows that, at large
$P_0$, the leading term of a propagator is $1/P_0^2$ times its
first moment $\int \! dx \, x \rho $.

The statement (b) is ultimately justified by evaluation. However,
the special assignment of the ${\cal L}$-term can be understood
immediately. At hard inner momentum the propagator $\Delta_t^-$
in ${\cal M}_1$ turns into the bare one. This is $\Delta_0$ and
cancels the extra factor $P^2$. Now, the remaining sum is easily
evaluated by means of (\ref{5singsum}) and (\ref{proptab}) to
give $-{\cal K} -{\cal L}/3$.

The statements (c) and (d) lead into further analysis. There is
a next step in common to all four expressions (see (\ref{rot})
below), if we are able to get rid of the extra $P$'s in
${\cal M}_1$ and ${\cal M}_3$. This is achieved by introducing
temporarily two more propagators:
\def\ms{\phantom{aaa} \vspace{.14cm} \\ }
\bea  \label{moreprop}
\hspace{-.4cm} \begin{array}{*{4}{c@{\;\;}}}
\mbox{propagator} \quad & \mbox{spectral density} \; \rho
            & \mbox{relation} & \quad \psi (p)  \ms
 \Omega_\theta & \rho_\theta \; (x^2-p^2) & \dis \Omega_\theta
              = P^2 \Delta_\theta - {2 \over 5} m^2 p^2
            &  \quad \dis - {8 \over 5} m^2 p^2 \ms
 \Lambda_t  & \rho_t \; (x^2-p^2)^2  & \Lambda_t = P^2 \Omega_t - m^2
            & \quad  m^2  \;\;\; .
\end{array}  \eea
With the 'relation'-column of (\ref{moreprop}) and with view to
Appendix D we obtain
\bea   \label{m145}
 {\cal M}_1 &=& {4 \over 15} {\cal L} - {1 \over 5} {\cal K} +
       {1 \over 6 m^2} \sum \Delta_t^- \Omega_\theta {1 \over p^2} \\
  {\cal M}_3 &=& {9 \over 2} {\cal K}
  - {3 \over 2 m^2} \sum \Lambda_t^- \Omega_\ell {1 \over p^2} \;\; .
\eea
Now all terms are either known (${\cal K}$, ${\cal L}$) or have the
desired form $\sum \Delta^- \Delta f$. Using the spectral
representation for both propagators and doing the frequency sum, one
is left with three integrations (over $p$ and two $x$). The following
formula reduces them to two. Let $A$ and $B$ be two of our
propagators (not necessarily different) and $\psi_A$, $\psi_B$ their
minus-first moments. Then:
\bea   \label{rot}
  \sum A^- B \, f(p) &=& \sum B^- A \, f(p) \;\, = \;\,
        T\int_P^3 f(p) \,\psi_A (p) \; \psi_B (p) \; +    \nonu \\
        & & \hspace{-2cm} {} + \; T\int_P^3 f(p) \int \! dx \,
           {1 \over x} \, \rho_A (x,p) \, {Q_0 \over x-Q_0} \,
           \bigg[ \, B(Q_0 -x,p) + \psi_B (p) \, \bigg] \;\; .
\eea
This formula may be derived in a straightforward manner. It is much
easier, however, to go in the backward direction. Inserting the
$\psi$-definitions (\ref{5singsum}) as well as the spectral
representation (\ref{5spect}) of $B$ to the right of (\ref{rot}) and
symmetrizing with respect to $x$, one arrives at the conclusion that
\be   \label{s}
  \sum_{P_0} \, {1 \over P_0^2 - x^2} \; {1 \over (P_0-Q_0)^2 -y^2}
  \; = \; {T \over x^2 y^2 } \, \left( 1 - {1 \over 2} \, S_x \,
          { Q_0 \, (Q_0 +x) \over (Q_0+x)^2 -y^2}   \right)
\ee
might have been used for the frequency sum on the left of
(\ref{rot}). $S_x f(x) \equiv f(x)+f(-x)$. If $x$ and $y$ are soft,
(\ref{s}) is indeed corrrect and is valid in the same sense as
(\ref{5frsum}). If $A \neq B$, (\ref{rot}) may be used in two
versions. In such a case we favour the transversal density $\rho_t$
to appear on the r.h.s., because it is slightly more convenient in
the numerical procedure.

Using the spectral representation of $B$ and taking (\ref{rot})
at $Q_0=m+i\varepsilon$, the imaginary part of (\ref{rot}) is
easily obtained as
\be  \label{imag}
   \Im m \,\, \sum A^- B \, f(p)  =
   {mT \over 2 \pi} \int_0^\infty \! dp \, f(p) \, p^2
   \int \! dx {1 \over x(x-m)} \;
   \rho_A (x,p) \rho_B(x-m,p) \;\; .
\ee
Using this formula for (\ref{magni}) one arrives at precisely the
analytical expression for the damping rate. The latter is equation
(25) in \cite{6.6}. Note that it was obtained there in Coulomb gauge.
Our longitudinal spectral density $\rho_\ell$ is $p^2/(p^2-x^2)$
times that of ref. \cite{6.6}.

We return to the real part. The analytical continuation
$Q_0 \rightarrow m + i \varepsilon $ was so far carried through in
trivial expressions only. However, in (\ref{rot}) this continuation
requires more care. Any propagator $B$ can be expressed by
$\Delta_\ell$ or $\Delta_t$. The functions $\P$ in their denominator
get an imaginary part. The real parts of $\Delta$ include this
Landau damping, see (\ref{Brea1}) and (\ref{Brea2}), and are
denoted in Appendix B by $\Delta^r (x,p)$. Note further that the
denominator $x-Q_0$ in (\ref{rot}) is harmless because at $x=Q_0$
the square bracket vanishes too: $\psi_B (p) = - B(0,p)$. To
summarize, after analytical continuation and when taking the real
part of (\ref{rot}), the propagator $\Delta$, which is $B$ or
occurs in $B$, has to be replaced by $\Delta^r(m-x,p)$. It includes
Landau damping if $p^2 > (m-x)^2$.

Consider ${\cal M}_2$ to see the above steps at work and to verify
the statement (c) as announced. Using (\ref{rot}), (\ref{Bmoment3}),
the table (\ref{proptab}) and the abbreviation
$\, t\equiv x-m \,$ we obtain
\bea  \label{M2}
  {\cal M}_2 &\! =\!& - 3 \, {m^2 T \over 2 \pi^2} \int_0^\infty
  \! dp \bigg( {\cal N}_\Delta + {\cal N}_\Omega  \bigg)
       \;\;\; \mbox{with} \;\;\;
  {\cal N}_\Delta \; = \; \int \! dx {1 \over x} \rho_t \left( 1 +
  {m\over t} \left[ 1 + p^2 \Delta_t^r (t,p) \right] \right) \nonu \\
\mbox{and} & & \!\!\!\! {\cal N}_\Omega \; = \; \int \! dx {1
   \over x} \rho_t \, {p^2-x^2 \over m t} \left[ 1 + (p^2-t^2)
   \Delta_t^r (t,p) \right]  \;\; .
\eea
The two terms ${\cal N}$ still correspond to the two terms in
(\ref{magni2}). To realize that both ${\cal N}$ are IR singular,
consider $p$ small, neglect $\rho_t^{\rm pole}$ and use the delta
function asymptotics $\rho_t^{\rm cut} / x \rightarrow \delta (x)
/ p^2 $, see (\ref{Bdelta}). With (\ref{Bsmallp}) one obtains
\be
  {\cal N}_\Delta \rightarrow  {5 \over 6 p^2 } \qquad , \qquad
  {\cal N}_\Omega \rightarrow  - {5 \over 6 p^2 } \;\; . \quad
\ee
Clearly each of these terms would make ${\cal M}_2$ divergent
in the infrared. But the two singularities cancel.

The above expression (\ref{M2}) also shows that by splitting off a
factor $mT$ of each ${\cal M}$, dimensionless quantities are obtained.
One may set $m=1$ in these quantities, whereafter the two integrations
run over dimensionless variables $p$ and $x$. This is the point where
further evaluation is relegated to the computer.

To summarize input and output, let $\; {\cal M}_0 \equiv 4
{\cal K}\;$ and $\;\eta_j \equiv 3 {\cal M}_j \, / \, m T\,$. Then:
\be   \label{etasum}
  \omega^2 \; = \; m^2 \left( 1 + \eta \, g \, \wu N \, \right)
  \qquad \mbox{with} \qquad \eta \; = \; \sum_{j=0}^4 \eta_j \;
  = \; -0.18 \lower 4.4pt\hbox{ $^2$} \quad , \qquad
\ee
where the result is composed of the following individual numbers:
\bea  \label{numbers}
\begin{array}{*{5}{c@{\,\,\,\quad}}}
    \eta_0 = + 0.55 \lower 4.4pt\hbox{$\, ^1$} &
    \eta_1 = - 0.13 \lower 4.4pt\hbox{$\, ^8$} &
    \eta_2 = - 0.25 \lower 4.4pt\hbox{$\, ^6$} &
    \eta_3 = - 0.12 \lower 4.4pt\hbox{$\, ^6$} &
    \eta_4 = - 0.21 \lower 4.4pt\hbox{$\, ^3$}    \quad .
\end{array}  \eea
Hence, all non-trivial contributions have the 'right' sign.
And the total is negative in accordance with the
intuitive picture given in the introduction. The third digits
are uncertain. Consequently, even the second digit in
(\ref{etasum}) can not be stated with conviction.
%
%
\let\dq=\thq \renewcommand{\theequation}{7.\dq}
\setcounter{equation}{0}

\parag {7. \ On the numerical problems }
This short last section is an attempt to list the pitfalls one
encounters in the numerical treatement of the twofold integral
in each of the four terms $\eta_j$. The stucture (\ref{M2}) is
typical. Both, a density $\rho$ and some expression linear in a
propagator $\Delta^r (1-x,p)$, develop their specialities over the
(dimensionless) $p$-$x$-plane. Consider for example
\bea
  \eta_4 &=& - {9 \wu 3 \over 16 \pi } + {9 \over 4 \pi^2}
     \int_0^\infty \! dp \int \! dx \, {1 \over x} \,
     \rho_\ell (x,p) \,\, (x^2-p^2) \,\, {\cal Q} (t,p) \nonu \\
  & & \mbox{with} \quad t \equiv x-1 \;\; , \;\; {\cal Q} (t,p)
      \,\; = \;\, {p^2 \over t} \, \left( {1 \over 3+p^2 } +
      {p^2-t^2 \over p^2} \Delta_\ell^r (t,p) \, \right)
\eea
and with $\rho_\ell$ and $\Delta_\ell^r$ from Appendix B. The delta
functions in the pole contribution of $\rho_\ell$ lead to single
integrals. In figure 3 they run along the dotted lines which start
from the points $x = \pm 1$ at the x-axis. In each step (to larger
$p$) the frequency $\omega_\ell (p)$ is determined numerically from
(B.9). On the whole left dotted line the propagator $\Delta^r_\ell$
is undamped. On the right it is damped.
      \fgrc    

The integration over the cut-part of $\rho_\ell$ is conveniently
done with the variables $a=p+x$, $b=p-x$. It runs over an undamped
pole of $\Delta_\ell^r\;$, see the dotted line inside the area
$p^2 >x^2$. We imbedded this pole in the sense $x / ( x^2 +
\varepsilon^2 ) $ and worked with a suitable variable step-width.
There is one more singular line at $a=1$ which separates the
damped/undamped regions of $\Delta_\ell^r\,$. With increasing $p$
these two singular lines approach each other exponentially. To
handle this speciality we separated a stripe around the
$a=1$-line and introduced a logarithmic variable
$\tau = - \ln (a-1)$ there. In passing, there were no problems at
the vertical line $x=1$ since the '' 0/0 '' on this line can be
avoided analytically.

Let a last warning stand at the end. In order to get all numerically
relevant pieces in the $p$-$x$-area, we had to run with $p$ up to
pretty high values: not 10, not 100, but 2500 !

%
%

\parag {8. \ Conclusions }
Braaten-Pisarski resummation is applied to calculate the real part
of the gluon plasmon frequency in the next-to-leading order.
Two of the three classes of contributions, which were predicted by
BP to be separate gauge independent sets within the order $O(g)$
of interest, do not reach this order. But this surprise of
a 'gauge independent number zero' is not in conflict with neither
prediction nor any principle. The gauge dependent terms in
the remaining class (soft one-loop diagrams in the effective
expansion) are explicitly shown to cancel out. Within the covariant
gauges (used therein) and within $O(g)$ we had no (true) IR-problem.
Also, the result is independent of the soft-hard threshold $q^\ast$.

The most laborious part has been the reformulation of the loop
contributions of subsection 5.2. Most probably, there is a shorter
and more elegant way of evaluation. However, once one is half way
inside the jungle, it is a hard decision to go back for only the
belief in a better world. For each contribution, we have
immediately restricted ourselves to the part $\Tr B \P $ and to the
limit $\vc q \rightarrow 0$. We were thus for instance unable to
check transversality \cite{transv} of the polarization function
to $O(g)$.

The next-to-leading order term has been obtained with a negative
coefficient. This minus sign is in accord with the intuitive
picture of a system whose longitudinal-electric mode becomes soft
with decreasing temperature and increasing coupling. The details
(although very special) support the general prospect of using the
known high temperature limit to understand QCD perturbatively
'from above'.

{\sl Permanently, while this work grew up (and ran into every
pitfall), there was a very enjoyable and helpful contact to Anton
Rebhan. Thanks to Fritjof Flechsig, who checked the imaginary parts,
an algebraic error could be eliminated (it invalidates the result
presented in the preprint foregoing this paper). I also acknowledge
encouraging discussions with Neven Bilic, Max Kreuzer, Rob Pisarski,
Martin Reuter and Uwe-Jens Wiese.}
%
%
\let\dq=\thq  \renewcommand{\theequation}{A.\dq}
\setcounter{equation}{0}
\def\aint{\int\!\!\!\!\!\!\!\>
          \lower 3pt\hbox{$\widehat{\hphantom{m}}$}\;}

\parag {Appendix \ A }
Here the various sums (\ref{zs}) of section 3 are evaluated. The
integration momenta read $K$ and $P$ and are considered hard in the
sense $q^\ast < k$ with $q^\ast = T \wu g $. But as far \nopagebreak
as no use is
made of this inequality we may play around with $q^\ast$. To start
with (and to introduce notations) consider $I_1$ and $J_0$ which we
combine in vector form:
\be \label{ji1}
\left\{ \, J_0\, ,\, I_1\, \right\} = \sum {
     \left\{ \, 1 \, , \, k^2 \, \right\} \over K^2 (K-Q)^2 }
     = {1 \over 2 \pi^2 } \int_{q^\ast}^\infty \!\! dk \;
       \left\{ \, k^2\, , k^4\, \right\} \; T \sum_n
       {1 \over K^2 (K-Q)^2 } \;\; .
\ee
Note that $K=(i\omega_n, \vc k )$ but $Q = (i\omega_{n^\prime},
\vc 0 )$ with $n^\prime$ an outer index. To do the frequency sum we
use the formula \cite{LW}
\be \label{frsum}
 T \sum_n F(i \omega_n ) = {1 \over 2 \pi i }
  \aint \! d\Omega \; F(\Omega) - {1 \over 2 \pi i }
  \oint \! d\Omega \; n(\Omega) S_\Omega F(\Omega )
  \equiv \; \aint + \oint \; \approx \; \oint \;\; ,
\ee
where $S$ is an operator which symmetrizes:
$S_\Omega F(\Omega ) = F(\Omega ) + F(-\Omega ) $. The intergal with
arrow runs along the imaginary axis in the $\Omega$-plane, that with
circle surrounds the right half plane counterclockwise. As is
indicated to the right of (\ref{frsum}) and reasoned in the main text,
we omit the temperature-independent arrowed integral in this section.
In (\ref{ji1}) the poles surrounded in the right-half plane lie at
$K_0=k$ and at $K_0=k+Q_0$. Note that $n(k+Q_0)=n(k)$. One obtains
\[
  T \sum_n {1 \over K^2 (K-Q)^2 } = - { n(k) \over k } \,
  {1 \over Q_0} \left( {1 \over Q_0 -2k } + {1 \over Q_0 +2k}
  \right)   \nonu
\]
and thus
\be \label{ji2}
  \left\{ \, J_0\, ,\, I_1\,\right\} = { 1 \over \pi ^2}
  \int_{q^\ast}^\infty \! dk \; k \, n(k) \,
  { \left\{ \, 1 \, , \, k^2 \,\right\} \over 4k^2 -Q_0^2 } \;\; .
\ee
For hard $k$ the $-Q_0^2$ in the denominator can be omitted, of
course. But $J_0$ also occurs in the 1-loop-hard analysis, where the
'strictly hard' part is to be subtracted. In the hard sense we have
thus obtained $I_1=T^2/24$. Note the relation $2I_1=-I_0$. We may see
of how this relation derives from (\ref{ji1}). Compared to $I_0$, the
factor $\vc k ^2 /(K-Q)^2$ introduces an additional pole (factor 2).
The extra propagator combines with the first as $-1/4k^2$. This gives
$-1/2$. In essence, the same mechanism also applies to $Z_1$ and $Z_0$
giving $2Z_1=-Z_0$. This way we avoid examining $Z_0$ here.

The sums $J_1$ and $J_0^\prime$ contain a $K^4$ in the denominator.
Surrounding $1/(K_0-k)^2$ leads also to a derivative of $n(k)$, which
one gets rid through integration by parts. The result is
\be \label{jj}
 J_1 = - {3 \over 4} J_0 \qquad {\rm and} \qquad  J_0^\prime =
       J_0^{\,\; Q_0 \rightarrow 0 }
       + {1 \over 4 \pi ^2} n(q^\ast ) \;\; .
\ee
Analytical continuation of the discrete $Q_0$-values amounts to
$Q_0 \rightarrow \omega + i \epsilon$. The 3/4-relation in (\ref{jj})
then holds true also for the imaginary parts.   \nopagebreak

We now turn to the doubly sums $Z_1$ and $Z_2$. In a first step
each sum is made explicit by using (\ref{ji1}) and (\ref{frsum}),
performing the two operations $S$ and marking the poles in the $P_0$
right-half plane. After the $P_0$-integration is done there remain
terms involving $n(\varpi )$ where $\varpi \equiv \vert \, \vc p
- \vc k \vert\,$. But under the spatial $P$-integral (which is
$\int_P^3 \equiv (2\pi)^{-3} \int d^3p$) the substitution $\vc p
\rightarrow \vc k - \vc p$ interchanges $\varpi $ with $p$. This way
we obtain
\bea \label{z1}
Z_{1,\, 2} &=&\int_K^3 \int_P^3
      \left\{ \, k^2\, ,\, {p^2 + \varpi^2 \over 2}\,\right\}
         {1 \over 2 \pi i } \oint dK_0 n(K_0) {1 \over K_0^2-k^2 }
         \left( {1 \over (K_0-Q_0)^2 -k^2 } \right. \nonu \\
& &  + \left. {1 \over (K_0+Q_0)^2 - k^2 } \right) {n(p) \over p}
  \left( {1 \over (K_0-p)^2 -\varpi^2 }
  + {1 \over (K_0+p)^2 -\varpi^2 } \right) \;\; .
\eea
With view to the $K_0$-integration, poles in the right-half plane
come from the denominators $K_0-k$, $K \pm Q_0 -k$, $K_0-p-\varpi $,
but from $K_0-p+\varpi$ and $K_0+p-\varpi$ only if $p>\varpi$ or
$p<\varpi$, respectively. The (lengthy) result will thus involve the
step functions $\theta (p-\varpi )$ and $\theta (\varpi -p)$.

All $Q_0$-dependence is in various denominators, and these get
$i \epsilon$ terms by analytical continuation. Pairs
of the corresponding delta-functions could contribute to the real
part. We find, however, that any two delta peaks either do not cross
or they force $k$ (or $p$) down to $O(\omega )$. Thus, and thereby
explicitly referring to $q^\ast < k$, we may omit all the $i \epsilon$
at all. Note that this no-contribution of delta pairs can be made
responsible for the non-$O(g)$ of the 2-loop diagrams.

In the following, $\omega$ stands in place of an earlier $Q_0$. After
the $K_0$-integral is done there are again unwanted $\varpi$ in some
Bose function arguments. This time we exploit the interchange of
$\varpi$ with $k$ under the corresponding substituion in the spatial
$K$-integral. To write down the next $Z$-version we need appropriate
notations,
\bea \label{nr}
N_\pm &=& {1\over (p\pm k)^2 -\varpi^2 } \; = \; {1 \over 2pk} \;
    {1\over u\pm 1} \quad , \quad u \equiv \cos (\vartheta ) \nonu \\
R_\pm &=& {1 \over (\omega \pm p + k )^2 -\varpi^2 } \quad , \quad
      R_{-}^{-} = R_{-}^{\;\, \omega \rightarrow - \omega }  \nonu \\
R_1 &=& {1 \over \omega^2 + 2k\omega + 2p(k - p) (1+u) } \quad ,
        \quad R_2 = R_1^{\; p \rightarrow -p}  \;\; ,
\eea
where $\vartheta$ is the angle between $\vc k$ and $\vc p$.
We may then write $Z_{1,\, 2}= S_\omega \; Z^{+}_{\, 1,\, 2} $ with
\bea \label{z2}
 Z^+_{\, 1,\, 2} &=& \int_K^3 \int_P^3  \left\{ \, k^2 \, ,
      \, {p^2+\varpi^2 \over 2} \, \right\} \, {n(p) \, n(k)
      \over 2 p k} \; {1 \over \omega (\omega + 2k) } \;
    \left( N_+ + N_- + R_+ + R_- \right) \nonu \\
& & + \int_K^3 \int_P^3 \left\{ \, \varpi^2 \, , \,
      {p^2+k^2 \over 2} \, \right\} {n(p) \over p }
      \left[ {n(p+k) \over 2k} N_+ R_+ \right. \nonu \\
& & - \left. \theta (p-k) {n(p-k) \over 2k} N_-R_-^-
      + \theta (k-p) {n(k-p) \over 2k} N_-R_-  \right]  \;\; .
\eea
The unwanted step functions can be recombined. To show this we
restrict to the relevant pieces of the twofold integral over the
square bracket,
\be \label{a0}
A = \int_0^\infty \! dk \, k^2 \int_{-1}^1 \! du \,
    \left\{ \, \varpi^2 \, , \, {p^2+k^2 \over 2} \, \right\} \;
    \bigg[ \; \ldots \; \bigg]
    \;\;\; \equiv A_1 + A_2 + A_3  \;\; ,
\ee
where $A_j$ refers to the $j$-th term in the square bracket. In
$A_1$ we substitute $k \rightarrow k-p$. This gives
\be \label{a1}
 A_1 = \int_p^\infty \! dk \int_{-1}^1\! du \; \left\{ \,
       k^2 + 2p(p-k)(1+u) \, ,\, {p^2+(k-p)^2 \over 2}\,\right\} \;
       {n(k)\over 4p(1+u)} R_1 \;\; .
\ee
In $A_2$ we substitute $k \rightarrow p-k$ followed by
$u \rightarrow -u$. This leads to just the above expression for $A_1$
except that the integration limits now are $0$ and $p$. $A_1$ and
$A_2$ thus combine to one integral from $0$ to $\infty$. In $A_3$ the
step function sets a lower limit at $p$. But this turns to $0$ aswell
by $k \rightarrow k+p$ (followed by $u \rightarrow -u$):
\be \label{a3}
A_3 = - \int_0^\infty \! dk \int_{-1}^1\! du\,\left\{ \, k^2
      + 2p(k+p)(1+u) \, ,\, {p^2+(k+p)^2 \over 2}\,\right\} \;
      {n(k)\over 4p(1+u)} R_2  \;\; .
\ee
The step functions have gone, and wanted arguments enter the Bose
functions. But we become aware that each integral $A_j$ diverges
when the angular integration runs down to $u=-1$. We show that these
singluarities cancel (in the case at hand, but not in $Z_2^\prime$).
Let us first summarise the form of $Z_{1,\, 2}^+$ so far reached:
\bea \label{z3}
 Z_{1,\, 2}^+ &=&  {1 \over 32 \pi ^4} \int_0^\infty \! dp \, n(p)
    \int_0^\infty \! dk \, n(k) \int_{-1}^1 \! du  \left( \, \left\{
  \, k^2\, ,\, {p^2+\varpi^2 \over 2}\,\right\} \, {2pk \over \omega
  (\omega +2k)} \,  \left[ N_+ + N_-  \right. \right.  \nonu \\
& & \left.\left.\hspace{-1.4cm} {} + R_+ + R_- \right]
    + \left[ \, \left\{ \, {k^2\over 1+u} + 2p(p-k) \; ,\;
      {p^2+(k-p)^2 \over 2\, (1+u)} \,\right\} \, R_1 \,
    - \;\mbox{ditto}_{p \rightarrow -p} \right] \right) \;\; .
\eea
If we decompose into partial fractions,
\[
 {1 \over 1+u} R_1 = {1 \over \omega (\omega +2k) } \left(
 {1 \over 1+u } + 2p(p-k) R_1 \right) \;\; ,
\]
the collinear singularity in $Z_1^+$ is removed immediately. Note that
$N_+ + N_- \rightarrow N_+ -N_+ =0$ by $u \rightarrow -u$. In $Z_2^+$
the crucial terms add up as
\[
   {- pk \over \omega (\omega +2k) } \left( {u \over 1+u} -
   {u \over 1-u} + {2 \over 1+u } \right)
   \rightarrow {- 2pk \over \omega (\omega +2k)} \;\; .
\]
The big round bracket in (\ref{z3}) now reads
\bea \label{bk} \hspace{-.8cm}
\bigg( \ldots \bigg) &=& {1 \over \omega (\omega +2k) } \left\{ \;
    2k^2 (Q_+ + Q_- ) + 2 (\omega +k)^2 (Q_1 - Q_2 ) \; , \;
    -4pk  \right. \nonu \\
& & + \left. \left(   \left[ p^2 + (p+k+\omega )^2 \right] Q_+
                     + \left[ p^2 + (p-k)^2 \right] Q_1
    - \mbox{ditto}_{p \rightarrow -p} \right) \;   \right\} \;\; , \\
& & \mbox{where} \;\; Q_\pm = pkR_\pm \;\; \mbox{and} \quad
                    Q_1 = p (p-k) R_1  \;\; , \nonu
\eea
and is ready to be integrated over the relative angle. For the
resulting logarithms we write $\ln (x)$ but mean $\ln ( \vert x
\vert )$. At this point we remember that evaluation at hard-hard
momenta is sufficient. We thus reintroduce $q^\ast$ as lower limits
in (\ref{z3}) and expand in powers of the soft $\omega$. For example:
\[
  { 1 \over \omega (\omega +2k )} = - {1 \over 4k^2 } + {1 \over
   2k\omega} + O(\omega) \quad , \qquad {(\omega +k)^2 \over \omega
   (\omega +2k)} = {3 \over 4} + {k \over 2\omega } + O(\omega )
\]
and
\bea  \label{iq+}  \hspace{-.8cm}
  2 \int_{-1}^1 \! du \; Q_+ &=& \ln \left( {4pk + 2(p-k)\omega +
  \omega^2 \over \omega \left[ 2p+2k+\omega \right] } \right) \nonu \\
&=& -\ln (\omega) + \ln \left( {2pk \over p+k} \right) +\omega
    {p+k \over 2pk} - \omega {1 \over 2p+2k } + O(\omega^2) \;\; .
\eea
Since $Z^+$ will be operated with $S_\omega$, only terms even in
$\omega$ need be retained. Then, at the end, $S_\omega$ amounts to
a factor of 2. Working this way we obtain the hard-hard part
\be  \label{ibk}
\int_{-1}^1 \! du \; \bigg( \ldots \bigg) = \left\{ \;
     \ln \left( {p+k \over p-k} \right) \; , \; {2p \over k} -
     {1 \over 2}  \ln \left( {p+k \over p-k} \right) + {p \over 2k}
     \ln \left( {k^2 \over p^2-k^2} \right) \; \right\} \;\; .
\ee
The first term in the second component leads to decoupled integrals,
which are easily identified with the hard parts of $I_0$ and $J_0$.
The result for $Z_{1,\, 2}$, as given in (\ref{z12}) in the main
text, is now obtained.

The sums $Z_1^\prime$ and $Z_2^\prime$ need not really be calculated,
since they cancel between different diagrams. We like to show,
however, how the collinear singularity looks like. We may start
with the expression (\ref{z1}) taken at $Q_0=0$. Due to the square
singularity $1/(K_0-k)^2$ the further calculation is slightly
different from the above. But (\ref{a0}) to (\ref{a3}) may be used
again, except that now $\omega=0$. At the level of (\ref{z3}) the
result is
\bea \label{zp0}
 Z_{1,\, 2}^\prime &=& {1 \over 16 \pi ^4} \int_0^\infty \! dp\, n(p)
    \int_0^\infty \! dk \, n(k) \int_{-1}^1 \! du  \left( \, \left\{
  \, k^2\, ,\, {p^2+\varpi^2 \over 2}\,\right\} \cdot \right. \nonu \\
& & \left[ \, \left( p {n^\prime (k) \over n(k) } -{p \over k} \right)
    (N_+ + N_- ) - 2p(p+k) N_+^2 + 2p(p-k) N_-^2 \, \right]  \nonu \\
& & \left. {} + \left[ \, \left\{ \, k^2 \, , \, {p^2+(p-k)^2 \over 2}
   \, \right\} {1 \over 1+u } N_1 - \, \mbox{ditto}_{p \rightarrow -p}
   \; \right] \;\; \right) \;\; ,  \\
& &  \mbox{where} \;\; N_1 = R_1^{\; \omega =0}
     = {-1 \over 2p(p-k)} \, {1 \over 1+u}  \;\; . \nonu
\eea
The second square bracket agrees with that of (\ref{z3}) if taken at
$\omega =0$ (which allowed the second term of the first
component to cancel). Consider the first component of (\ref{zp0}).
Making explicit all denominators containing $u$ and using
$u \rightarrow -u$, the collinear singularity can be packed up in
a separate factor:
\be  \label{zp1}
 Z_1^\prime = {1 \over 16\pi^4} \int_{-1}^1 \! du \left(
     {1 \over 1+u} \right) ^2 \, \int_0^\infty \! dp \, n(p)
     \int_0^\infty \! dk \, n(k) \, { kp \over k^2-p^2} \, = 0 \;\; .
\ee
The whole term vanishes because the $p$-$k$-integral runs over
a function which is antisymmetric under an interchange of $p$ with
$k$. We turn to the second component and treat it in a similar
manner. Singular terms can again be localized, but now their
prefactors remain non-zero:
\bea  \label{zp2}
 Z_2^\prime &=& {1 \over 16\pi^4}  \int_0^\infty \! dp \, n(p) \, p
       \int_0^\infty \! dk \, \Bigg( -2 n^\prime (k) + n^\prime (k)
       \int_{-1}^1 \! du {1 \over 1+u} \hspace{1.8cm}   \nonu \\
& & \hskip 4.4cm {} + { 2 \over k} n(k) - {2 \over k} n(k) \int_{-1}^1
    \! du \left[ {1 \over 1+u } \right] ^2 \;\; \Bigg) \;\; .
\eea

The last object to be considered is $Z_0^\prime$. After all, it is a
rather simple sum and becomes zero through $N_+ + N_- \rightarrow 0$.
To end up, we note that this is in accord with the relation
$\, 2 Z_1^\prime = - Z_0^\prime\, $ as both sides vanish.
\vspace{-.3cm}
%
%
\let\dq=\thq  \renewcommand{\theequation}{B.\dq}
\setcounter{equation}{0}

\parag {Appendix \ B}
Here properties of the propagators $\Delta_t$ and $\Delta_\ell$ are
detailed. These propagators are defined in (\ref{gmn}) and related
by (\ref{deltas}) to the polarization functions
$\P_t = {1 \over 2} \Tr A\Pi$ and $\P_\ell = \Tr B \P$ at
one-loop order. While here we merely only list the known facts
on $\P$, $\Delta$ and their spectral densities $\rho$, they
aquire a unique notation. From (\ref{piyy}) to (\ref{l+t})
and with $P_0$ a complex variable still already apart from the
imaginary axis:
\bea  \label{Bpp1}
 \P _t \; (P) \; = \,\; {3 \over 2} m^2 g \left( P_0 \over p \right)
         \quad & , & \quad
 \P _\ell \; (P) \; = \,\; 3 m^2 \left[ \, 1 - g \left( {P_0 \over p}
     \right) \,\right]     \\  \label{Bpp2}
 \mbox{with} \quad  g(z) & = &  z^2 - {1 \over 2} z \left( z^2 - 1
   \right) \ln \left( { z+1 \over z-1 } \right) \\
 g(z) & = & {2 \over 3} + {2 \over 15 z^2 } + {2 \over 35 z^4 } + \;
 \ldots  \qquad (\, z \rightarrow \infty\, ) \;\; . \qquad \nonu
\eea
If $P_0$ approaches the real x-axis, one derives from
(\ref{Bpp1}), (\ref{Bpp2}) that
\be   \label{Bim}
    \Im m \,\; \P_t (x + i \varepsilon , p) = - {3 \pi \over 4}
    \, \xi \, \eta \; \theta (p^2-x^2) \quad , \quad \xi \equiv
    {x \over p} \quad , \quad \eta \equiv 1 - \xi^2 \;\; ,
\ee
The corresponding real part is (\ref{Bpp1}), (\ref{Bpp2}) with the
logarithm taken at the absolute value of its argument. For the
imaginary part of $\P_\ell$ one may use the identity $\P_\ell + 2
\P_t = 3 m^2$. From (\ref{deltas}) and (\ref{Bim}) and by anticipating
the notations of (\ref{Brho2}) we obtain
\bea   \label{Brea1}
    \Re e \,\Delta_t (x+i\varepsilon , p ) & = &
       {4 \over m^2} \, {- D_t \over D_t^2 +
                \theta (p^2-x^2) \, C_t^2}
       \;\; \equiv \; \Delta_t^r (x,p)  \;\; ,  \\  \label{Brea2}
    \Re e \,\Delta_\ell (x+i\varepsilon , p ) & = &
       {2 \over m^2 \,\eta} \, {- D_\ell \over D_\ell^2 +
                \theta (p^2-x^2) \, C_\ell^2}
       \;\; \equiv \; \Delta_\ell^r (x,p) \;\; .
\eea
(\ref{Brea1}) and (\ref{Brea2}) are even functions of x. For a
special purpose in section 6 we add the small-$p$ behaviour
\be   \label{Bsmallp}
  \Delta_t^r (m,p) \; = \; - {5 \over 6 \, p^2} \; + \; O(1) \;\; .
\ee

The spectral densities \cite{pispect} $\rho_t$ and $\rho_\ell$ have
a common structure :
\be  \label{Brho1}
   \rho = \rho^{\rm pole} + \rho^{\rm cut} \; , \,\;
   \rho^{\rm pole} = r\, \delta (x-\omega) - r\, \delta (x+\omega )
   \; ,\,\; \rho^{\rm cut} = \theta (q^2-x^2 ) \, {1 \over m^2 }
   { N \over D^2 + C^2 }  \; , \;
\ee
\vspace{-.5cm} \def\loga{ \ln \left( { 1+\xi \over 1-\xi } \right) }
\bea  \label{Brho2}
  \mbox{with} \qquad \qquad
  r_t = { \omega_t \,\, (\omega_t^2 -p^2 ) \over 3 m^2 \omega_t^2
        - (\omega_t^2 -p^2)^2 }  & , &
  r_\ell = { \omega_\ell \over 3 m^2 - \omega_\ell^2 + p^2 }
           \;\; , \nonu \\
  N_t = 12 \xi \eta & , & N_\ell = - 6 \, { \xi \over \eta }
           \;\; , \nonu \\
  C_t = 3 \pi \xi \eta  & , & C_\ell = 3 \pi \xi \;\; , \nonu \\
  D_t = 4 {p^2 \over m^2} \eta + 6 \xi^2 + 3 \xi \eta \loga & , &
  D_\ell = 2 {p^2 \over m^2} + 6 - 3 \xi \loga \;\; ,
\eea
The frequencies $\omega_t$, $\omega_\ell$ are the positive solutions
($\neq p$) of
\be   \label{Bomega}
\omega^2 = p^2+\P_j (\omega , p)\qquad (\; j = t\, , \,\ell\; )\;\; .
\ee
They are obtained by solving (\ref{Bomega}) numerically.

The most important moments \cite{BY,ich}
of the densities $\rho$ are
\bea  \label{Bmoment1}
  n=1  & & \quad \int \! dx \; x \; \rho_j (x, p) = 1
     \qquad \quad \qquad \qquad \;\; (\; j= t,\, \ell \;)\quad , \\
  \label{Bmoment2}
  n=3  & & \quad \int dx \; x^3 \; \rho_j (x, p) = m^2 + p^2
           \qquad \qquad  (\; j=t, \, \ell \;)\quad , \\
  \label{Bmoment3}
  n=-1 & & \quad \int \! dx \; {1 \over x } \; \rho_t (x,p)
     = {1 \over p^2 }  \;\;\; , \;\;\; \int \! dx \; {1 \over x } \;
     \rho_\ell (x,p) = {1 \over 3 m^2 + p^2} \;\;\; . \qquad
\eea

All the above details of the spectral densities are particularly
important in the course of the numerical evaluation of the
'magnificient seven' of section 6. Note that the two cut-parts of
$\rho$ behave quite different. Consider $p$ fixed and let $x$
approach the borders $\pm p$ of the interval set by the step function.
Then $\rho_\ell^{\rm cut}$ runs to $-\infty$ due to the prefactor
$1/\eta$. Only the squared logarithm in $D_\ell$ keeps the integal
in e.g. (\ref{Bmoment1}) finite. On the other hand, the cut part
of the transversal density has no such prefactor. If $p <\!\! < 1$,
this density is concentrated around $x=0$. With
$v \equiv { 4 \over 3 \pi m^2 }$ one obtains
\be  \label{Bdelta}
 {1 \over x} \rho_t^{\rm cut} (x,p)\;\rightarrow \;\theta (p^2-x^2) \;
 {1 \over \pi} \; { v p \over x^2 + v^2 p^6 } \;\rightarrow \;
 {1 \over p^2} \, \delta (x) \qquad (\, p \rightarrow 0 \, ) \;\; .
\ee
Note that this formula trivially leads to the minus-first moment
(\ref{Bmoment3}). But in the other two moments (\ref{Bmoment1}),
(\ref{Bmoment2}), the pole-contribution dominates at small $p$.
\newpage
%
%
\let\dq=\thq  \renewcommand{\theequation}{C.\dq}
\setcounter{equation}{0}

\parag {Appendix \ C }
Here we comment upon the treatment of the hard-hard double sums,
which occur in the loop diagram of subsection 5.2. Then the
resulting single sums (the 'greek sums') are collected. A typical and
sufficient general example is (\ref{waw}). The denominator is already
factorized. Thus the only task is rewriting the numerator
appropriately. As (\ref{waw}) shows, this leads to products of the
following three single sums:
\bea
 \sum_K {K_0^2 \over N } \vc k
      &=& \sigma \, {\vc p \over p^2 } \\
 \sum_K {K_0 \over N } \vc k \circ \vc k
      &=& \beta \, {1 \over 2} \left( 1 - {\vc p \circ \vc p \over
      p^2} \right) + \gamma \, {\vc p \circ \vc p \over p^2} \\
 \sum_K {k_i k_j k_\ell \over N } &=& \rho \, {1 \over 2} \left[
   \left(\delta_{ij} - {p_i p_j \over p^2} \right) {p_\ell \over p^2}
   + \left(\delta_{j \ell } - {p_j p_\ell \over p^2} \right) {p_i
   \over p^2}  \;\, + \right. \nonu \\
  & & \left. {} + \left(\delta_{i \ell } - {p_i p_\ell \over p^2}
      \right) {p_j \over p^2} \, \right] + \zeta \, {p_i p_j p_\ell
      \over p^4 } \;\; ,
\eea
where $N \equiv K^2 (K-Q)^2 (K-P)^2$ ($= N_K$ in the main text). The
form of the right-hand sides is dictated by symmetry. Note that $\vc q
=0$. Thus, the only direction, which the $K$-sums 'know' of, is that
of $\vc p$. The coefficients $\sigma$, $\beta$, $\gamma$, $\rho$ and
$\zeta$ are determined by taking traces and by multiplications with
$\vc p$. The results are given in (C.9) to (C.13) below.

There occur ten such 'greek sums' in the main text. Irrespecitve of
several relations between these sums, we shall list them all as in a
table of integrals. Their evaluation towards hard integration momentum
$K$ is rather familar and not given here. Note that the two outer
momenta $Q$ and $P$ are soft. The shorthand notations
$\P_\ell = \P_\ell (P)\,$, $\P_\ell^- = \P_\ell (P-Q)\,$,
$\Delta_0 = 1 / P^2 $ and $\Delta_0^- = 1 / (P-Q)^2$ are also used
in the main text. But $\overline{P_0} \equiv Q_0 - P_0 \,$,
$[-] \equiv [ k^2 - (\vc k \vc p )^2 / p^2 ]$ and
$\phi \equiv 8 g^2 N$ are special to the following table.
Some of the results simplify by using $b \equiv m^2 -\P_\ell \,$.

\hspace{2.5cm} definition \hspace{3cm} result $\;$( leading terms )
\bea
\mu = \sum {KP-KQ \over N } \; [-]
   &\quad ,\quad & \phi \, \mu = \P_\ell - m^2  = -b  \hspace{3cm} \\
\tau = \sum {(KP)\; (KQ-KP) \over N } \; {\vc k \vc p \over p^2}
        & , & \phi \, \tau = m^2                \\
\vartheta = \sum {KP \over N } \; [-]
        & , & \phi \, \vartheta = \P_\ell^- - m^2  = -b^-   \\
\varphi = \sum {KP \over N } \; {(\vc k \vc p )^2 \over p^2}
        & , & \phi \, \varphi = m^2 - \overline{P_0}^2 \,
              \Delta_0^- \, \P_\ell^- \\
\lambda = \sum {K_0 \; (KP) \; (\vc k \vc p ) \over N }
        & , & \phi \, \lambda = p^2 \, \overline{P_0} \,
              \Delta_0^- \, \P_\ell^-  \;\; .
\eea
The above five sums occur in the gauge dependent contributions
from the soft loop. The following five sums, which determine the
physical soft loop contributions, are either symmetric (first three)
or antisymmetric (last two) under the shift $P \rightarrow Q-P$:
\bea
\rho = \sum { \vc k \vc p \over N } \; [-]
      & \quad , \quad & \phi \, Q_0 \, \rho = Q_0 \, m^2
          - P_0 \, \P_\ell - \overline{P_0} \, \P_\ell^-
     = P_0 \, b + \overline{P_0} \, b^- \qquad \\
\sigma = \sum {K_0^2 \; (\vc k \vc p ) \over N }
      & , & \phi \, Q_0 \, \sigma = p^2 \, \Delta_0 \, P_0 \, \P_\ell
     + p^2 \, \Delta_0^- \, \overline{P_0} \, \P_\ell^-  \\
\zeta = \sum {1 \over N} \; {(\vc k \vc p )^3 \over p^2 }
      & , & \phi \, Q_0 \, \zeta = - Q_0 \, m^2 + \Delta_0 \, P_0^3 \,
            \P_\ell + \Delta_0^- \, \overline{P_0}^3 \, \P_\ell^-  \\
\beta = \sum {K_0 \over N } \; [-]
      & , & \phi \, Q_0 \, \beta = \P_\ell^- - \P_\ell = b - b^- \\
\gamma = \sum {K_0 \over N } \; { (\vc k \vc p )^2 \over p^2}
      & , & \phi \, Q_0 \, \gamma = \Delta_0 \, P_0^2 \, \P_\ell
              - \Delta_0^- \, \overline{P_0}^2 \, \P_\ell^- \;\; .
\eea
Relations between these sums are given in (\ref{grrel1}) and
(\ref{grrel2}) in the main text. They can be derived directly
from the definitions.
%
%
\let\dq=\thq  \renewcommand{\theequation}{D.\dq}
\setcounter{equation}{0}
\def\K{{\cal K}}  \def\L{{\cal L}}  \def\V{{\cal V}}

\parag {Appendix \ D }
Here the soft sums over single propagators are collected,
which occur in section 5. For their evaluation see
(\ref{5spect}) to (\ref{5L}) as well as the table (\ref{proptab}).
For the quantities ${\cal K}$, ${\cal L}$ and ${\cal V}$, the
results are formulated with, see also the main text at
(\ref{5L}), (\ref{5K}) and (\ref{5V}).
\bea
 \sum \Delta_\ell \; = \;
 \sum \left( \Delta_\ell^- - \Delta_\ell \right) {P_0 \over Q_0}
     & = & - \L  \hspace{3cm}     \\
 \sum \Delta_\ell \, {m^2 \over p^2} \; = \;
 \sum \left( \Delta_\ell^- - \Delta_\ell \right) {P_0 \over Q_0}
    \; {m^2 \over p^2} & = & - \K    \\
 \sum \Delta_0 \; = \; \sum \Delta_t \; = \;
 \sum \left( \Delta_t^- - \Delta_t \right) {P_0 \over Q_0}
     & = &  - 3 \K - \L    \\
 \sum \Gamma_\ell \, {1 \over p^2} \; = \;
 \sum \left( \Gamma_\ell^- - \Gamma_\ell \right) {P_0 \over Q_0} \;
      {1 \over p^2} \; = \;  {1 \over m^2 } \sum
      \Gamma_\ell^- \; {P_0^2 \over p^2} & = & 2 \K    \\
 {1 \over m^2} \sum \Gamma_\ell \; = \; 2 \L \quad , \quad
      {1 \over m^2} \sum \Gamma_t & = & -3\K -\L   \\
 \sum \Delta_t \, {m^2 \over p^2} \; = \;
 \sum \left( \Delta_t^- - \Delta_t \right) {P_0 \over Q_0} \;
     {m^2 \over p^2} & = &  - \V       \\
 \sum \Gamma_t \, {1 \over p^2} \; = \;
 \sum \left( \Gamma_t^- - \Gamma_t \right) {P_0 \over Q_0} \;
      {1 \over p^2}  & = & - \V     \\
 \sum \Gamma_\ell \; {P_0^2 \over p^2 } \; = \; \sum \Gamma_t \;
      {P_0^2 \over p^2 } & = & 0   \\
 \sum \Omega_t \; = \; 0 \quad , \quad
      {1 \over m^2} \sum \Omega_\ell & = & -3 \L
\eea
%
                    \renewcommand{\section}{\paragraph}

\end{document}